\documentclass{ptephy}
\renewcommand{\thesection}{\S\,\arabic{section}}

\begin{document}

\title{Hidden Nambu mechanics: \\
{\LARGE A variant formulation of Hamiltonian systems }}

\author{\name{\fname{Atsushi} \surname{Horikoshi}}{1,\ast}
and 
\name{\fname{Yoshiharu} \surname{Kawamura}}{2,\dag}
}

\address{\affil{1}{Department of Natural Sciences, Tokyo City University,
Tokyo 158-8557, Japan}
\affil{2}{Department of Physics, Shinshu University,
Matsumoto 390-8621, Japan}
${}^\ast$\!\! {\rm E-mail: horikosi@tcu.ac.jp} \\
${}^\dag$\!\! {\rm E-mail: haru@azusa.shinshu-u.ac.jp}
}

\begin{abstract}%
We propose a variant formulation of Hamiltonian systems
by the use of variables including redundant degrees of freedom.
We show that Hamiltonian systems can be described by extended dynamics 
whose master equation is the Nambu equation or its generalization.
Partition functions associated with the extended dynamics 
in many degrees of freedom systems are given.
Our formulation can also be applied to Hamiltonian systems 
with first class constraints. 
\end{abstract}

\subjectindex{A00, A30}

\maketitle
\section{Introduction}\label{Introduction}

In general, we have a choice of variables describing a physical system.
In most cases, we choose a set of variables 
whose number is same as the total number of degrees of freedom of the system
so as to minimize the number of equations of motion.
However, in some cases, it is quite useful 
to formulate the system by the use of variables including redundant ones.
A system with gauge symmetry offers a typical example.
To describe such a system, keeping the gauge symmetry manifest, 
we should employ a formulation that includes redundant variables. 
Although such a formulation is somewhat complicated, thanks to the symmetry,
we can clearly understand the important properties of the system 
such as conservation laws and form of interactions,
and can also calculate physical quantities in a systematic way \cite{W1,W2}.

Therefore, it is interesting to explore the general features of formulations 
including redundant degrees of freedom.
Here we base this on a principle (or brief) that
{\it physics should be independent of the choice of variables to describe it},
and make an attempt to formulate
Hamiltonian systems (systems of Hamiltonian dynamics)
in terms of new sets of variables including redundant ones.
What kind of dynamics describes the time evolution of the new variables?

Our strategy and conjecture are as follows.
Consider a Hamiltonian system described by
a canonical doublet $(q, p)$.
Take $N(\ge 3)$ variables $(x_1, \cdots, x_{N})$
that are functions of the canonical doublet,
and deal with them as fundamental variables to describe the system.
If they contain redundant variables, constraints 
between some variables must be induced.
To handle the constraints,
Dirac formalism \cite{D1,D2} provides a helpful perspective, 
where constraints with Lagrange multipliers are added to the original Hamiltonian.
The induced constraints play a similar role to the Hamiltonian.
As for the dynamics of $N$ variables,
Nambu mechanics \cite{N} is quite suggestive.
In Nambu mechanics,
fundamental variables form an $N$-plet, 
whose time evolution is generated by $N-1$ Hamiltonians
according to the Nambu equations. 
Combining the advantages of the two theories,
we conjecture that 
{\it there is a formulation whose master equation has a form of the Nambu equation 
or its generalization, 
where the Hamiltonians consist of the original one and the induced constraints.}

Nambu mechanics is a generalization of the Hamiltonian dynamics 
proposed by Nambu forty years ago \cite{N}.
In his formulation, the dynamics of an $N$-plet 
is given by the Nambu equation, 
which is defined by $N-1$ Hamiltonians and the Nambu bracket, 
a generalization of the Poisson bracket. 
The structure of Nambu mechanics is so elegant that many authors
have investigated its application.
However, the applications have been limited to particular systems
such as constrained systems, superintegrable systems, and hydrodynamic systems,
because Nambu systems (systems of Nambu mechanics)
should have multiple Hamiltonians or conserved quantities.
For example, researchers have studied 
how Nambu mechanics can be embedded into 
constrained Hamiltonian systems \cite{BF,CK,R,MS,KT,KT2}
or how constrained systems can be described in terms of Nambu mechanics \cite{LJ}.

In this article, we show that the structure of Nambu mechanics is, 
in general, hidden in systems of Hamiltonian dynamics.
That is, Hamiltonian systems can be described 
by Nambu mechanics or its generalization
by means of a change of variables 
from canonical doublets to multiplets.
Our formulation can be generalized to 
many degrees of freedom systems,
and the associated partition functions are given.
We also apply our formulation to systems with first class constraints.
Our approach can be regarded as a complementary one to 
the previous works \cite{BF,CK,R,MS,KT,KT2,LJ}.

The outline of this article is as follows.
In the next section, we give a formulation of Hamiltonian systems
using Nambu mechanics and its generalizations.
As an application, Hamiltonian systems with first class constraints 
are also formulated as Nambu systems in Sect. 3.
In the last section, we give conclusions and discussions on the direction of 
future work.
In Appendix A, we derive the Nambu equation from the least action principle.
In Appendix B, we show that a Nambu system of an $N$-plet
can be described by Nambu mechanics with an $N+r$-plet ($r\ge1$).

\section{Nambu systems hidden in Hamiltonian systems}
\label{Nambu systems hidden in Hamiltonian systems}

\subsection{Review}
\label{Review}
We begin with a brief review of Hamiltonian systems 
and Nambu systems \cite{N}.
A Hamiltonian system is a classical system described by 
a generalized coordinate $q=q(t)$ and its canonical conjugate momentum $p=p(t)$.
These variables satisfy the Hamilton's canonical equations of motion,
\begin{eqnarray}
\frac{d q}{dt} = \frac{\partial H}{\partial p} ~,~~ \frac{d p}{dt} 
= -\frac{\partial H}{\partial q}~,
\label{H-eq}
\end{eqnarray}
where $H=H(q, p)$ is the Hamiltonian of this system.
For any functions $A=A(q,p,t)$ and $B=B(q,p,t)$,
the Poisson bracket is defined by means of the 2-dimensional Jacobian,
\begin{eqnarray}
\{A, B\}_{\mbox{\tiny{PB}}} \equiv  
\frac{\partial (A,B)}{\partial (q,p)}=
\frac{\partial A}{\partial q}\frac{\partial B}{\partial p}
 - \frac{\partial A}{\partial p}\frac{\partial B}{\partial q}~.
\label{PB}
\end{eqnarray}
In terms of the Poisson bracket,
the Hamilton's canonical equation of motion for any function $f=f(p,q)$ 
can be written as
\begin{eqnarray}
\frac{d f}{dt} = \{f, H\}_{\mbox{\tiny{PB}}}~.
\label{H-eqf}
\end{eqnarray}

On the other hand, a Nambu system is a classical system described by a multiplet. 
As the most simple example, 
let us consider a Nambu system described by a triplet 
$x=x(t)$, $y=y(t)$, and $z=z(t)$.
These variables satisfy the Nambu equations
\begin{eqnarray}
\frac{d x}{dt} = \frac{\partial ({H}_1, {H}_2)}{\partial (y, z)}~,~~ 
\frac{d y}{dt} = \frac{\partial ({H}_1, {H}_2)}{\partial (z, x)}~,~~
\frac{d z}{dt} = \frac{\partial ({H}_1, {H}_2)}{\partial (x, y)}~,
\label{N-eq}
\end{eqnarray}
where ${H}_1(x, y, z)$ and ${H}_2(x, y, z)$ are $\lq\lq$Hamiltonians" of this system.
For any functions $A={A}(x, y, z, t)$, $B={B}(x, y, z, t)$, and $C={C}(x, y, z, t)$,
the Nambu bracket is defined by means of the 3-dimensional Jacobian,
\begin{eqnarray}
\{{A}, {B}, {C}\}_{\mbox{\tiny{NB}}} 
\equiv \frac{\partial ({A}, {B}, {C})}{\partial (x, y, z)}~.
\label{NB}
\end{eqnarray}
In terms of the Nambu bracket,
the Nambu equation for any function $f=f(x,y,z)$ 
can be written as
\begin{eqnarray}
\frac{d {f}}{dt} = \{{f}, {H}_1, {H}_2\}_{\mbox{\tiny{NB}}}~.
\label{N-eqf}
\end{eqnarray}

It is straightforward to extend the above formalism
to a system described by an $N$-plet $x_i$ $(i=1, 2, \cdots, N)$.
These variables satisfy the Nambu equations
\begin{eqnarray}
\frac{d x_i}{dt} =  \sum_{i_1, \cdots, i_{N-1}=1}^{N}
\varepsilon_{i i_1 \cdots i_{N-1}}
\frac{\partial {H}_1}{\partial x_{i_1}} \cdots 
\frac{\partial {H}_{N-1}}{\partial x_{i_{N-1}}}~,
\label{N-eq-N}
\end{eqnarray}
where ${H}_a={H}_a(x_1, x_2, \cdots, x_N)$ $(a=1, \cdots, N-1)$ are 
$\lq\lq$Hamiltonians" of this system
and $\varepsilon_{i i_1 \cdots i_{N-1}}$ is the $N$-dimensional Levi--Civita symbol, 
the antisymmetric tensor with $\varepsilon_{12 \cdots N} =1$.
For any functions 
${A}_{\alpha}={A}_{\alpha}(x_1, x_2, \cdots, x_N, t)$ $(\alpha=1, \cdots, N)$,
the Nambu bracket is defined by means of the $N$-dimensional Jacobian,
\begin{eqnarray}
\{{A}_1, {A}_2, \cdots, {A}_N\}_{\mbox{\tiny{NB}}} 
&\equiv& 
\frac{\partial ({A}_1, {A}_2, \cdots, {A}_N)}{\partial (x_1, x_2, \cdots, x_N)}
\nonumber \\
&=& \sum_{i_1, i_2, \cdots, i_N=1}^{N} \varepsilon_{i_1 i_2 \cdots i_N}
\frac{\partial {A}_1}{\partial x_{i_1}}\frac{\partial {A}_2}{\partial x_{i_2}}
 \cdots \frac{\partial {A}_{N}}{\partial x_{i_N}}~.
\label{NB-N}
\end{eqnarray}
In terms of the Nambu bracket,
the Nambu equation for any function ${f}={f}(x_1, x_2, \cdots, x_N)$
can be written as
\begin{eqnarray}
\frac{d {f}}{dt} = \{{f}, {H}_1, {H}_2, \cdots, {H}_{N-1}\}_{\mbox{\tiny{NB}}}~.
\label{N-eqf-N}
\end{eqnarray}

\subsection{Hidden Nambu structure}
\label{Hidden Nambu structure}

Here let us describe a Hamiltonian system with a canonical doublet $(q, p)$ 
by means of $N$ variables $x_i=x_i(q, p)$ $(i=1, \cdots, N)$.

\subsubsection{Formulation}
\label{Formulation}
\par
First we study the case with $N=2$, for completeness.
We assume that $x=x_1(q, p)$ and $y=x_2(q, p)$
satisfy $\{x, y\}_{\mbox{\tiny{PB}}} \ne 0$.
In this case, the equation for a function 
$\tilde{f}(x, y) = f(q, p)$
is written as 
\begin{eqnarray}
\frac{d \tilde{f}}{dt} = \frac{\partial(f, H)}{\partial(q, p)}
= \frac{\partial(\tilde{f}, \tilde{H})}{\partial(x, y)}
\frac{\partial(x, y)}{\partial(q, p)} 
= \frac{\partial(\tilde{f}, \tilde{H})}{\partial(x, y)} \{x, y\}_{\mbox{\tiny{PB}}}~,
\label{H-eq(N=2)}
\end{eqnarray}
where $\tilde{H}(x, y) = H(q, p)$.
If $\{x, y\}_{\mbox{\tiny{PB}}}=1$, the transformation $(q, p) \to (x, y)$ 
is the canonical transformation, and
$(x, y)$ are canonical variables.

Next we study the case with $N=3$.
We assume that variables $x=x_1(q, p)$, $y=x_2(q, p)$, and $z=x_3(q, p)$ 
satisfy at least two of the conditions
$\{x, y\}_{\mbox{\tiny{PB}}} \ne 0$, $\{y, z\}_{\mbox{\tiny{PB}}} \ne 0$,
and $\{z, x\}_{\mbox{\tiny{PB}}} \ne 0$.
In this case, the equation for a function 
$\tilde{f}(x, y, z) = f(q, p)$
is written as 
\begin{eqnarray}
\frac{d \tilde{f}}{dt} = \frac{\partial(f, H)}{\partial(q, p)} 
= \frac{\partial(\tilde{f}, \tilde{H})}{\partial(x, y)} \{x, y\}_{\mbox{\tiny{PB}}}
+ \frac{\partial(\tilde{f}, \tilde{H})}{\partial(y, z)} \{y, z\}_{\mbox{\tiny{PB}}}
+ \frac{\partial(\tilde{f}, \tilde{H})}{\partial(z, x)} \{z, x\}_{\mbox{\tiny{PB}}}~,
\label{H-eq(N=3)}
\end{eqnarray}
where $\tilde{H}(x, y, z) = H(q, p)$.
Note that $q$, $p$, and $H$ are, in general, 
not uniquely determined as functions of $x$, $y$, and $z$.

Introducing a function $\tilde{G}=\tilde{G}(x, y, z)$ that satisfies the conditions
\begin{eqnarray}
\frac{\partial \tilde{G}}{\partial x} = \frac{\partial(y, z)}{\partial(q, p)}~,~~
\frac{\partial \tilde{G}}{\partial y} = \frac{\partial(z, x)}{\partial(q, p)}~,~~
\frac{\partial \tilde{G}}{\partial z} = \frac{\partial(x, y)}{\partial(q, p)}~,
\label{xyzG}
\end{eqnarray}
Eq. (\ref{H-eq(N=3)}) is rewritten 
as the Nambu equation in the form of Eq. (\ref{N-eqf}),
\begin{eqnarray}
\frac{d \tilde{f}}{dt} = \{\tilde{f}, \tilde{H}, \tilde{G}\}_{\mbox{\tiny{NB}}}~,
\label{H-eqf(N=3)}
\end{eqnarray}
where we use the formula
\begin{eqnarray}
\frac{\partial (\tilde{A}, \tilde{B}, \tilde{C})}{\partial (x, y, z)}
= \frac{\partial (\tilde{A}, \tilde{B})}{\partial (x, y)}
\frac{\partial \tilde{C}}{\partial z}
+ \frac{\partial (\tilde{A}, \tilde{B})}{\partial (y, z)}
\frac{\partial \tilde{C}}{\partial x}
+ \frac{\partial (\tilde{A}, \tilde{B})}{\partial (z, x)}
\frac{\partial \tilde{C}}{\partial y}~.
\label{J(N=3)}
\end{eqnarray}
The conditions (\ref{xyzG}) are compactly expressed as
\begin{eqnarray}
\frac{\partial \tilde{G}}{\partial x_i} 
= \frac{1}{2} \sum_{j, k=1}^3 \varepsilon_{ijk}\{x_j, x_k\}_{\mbox{\tiny{PB}}}~~~
\mbox{or}~~~
\sum_{k=1}^3 \varepsilon_{ijk} \frac{\partial \tilde{G}}{\partial x_k} 
= \{x_i, x_j\}_{\mbox{\tiny{PB}}}~.
\label{xyzG2}
\end{eqnarray}
In Appendix A, the Nambu equations in the form of Eq. (\ref{N-eq}) 
are also derived from a Hamiltonian system with a canonical doublet $(q, p)$
using the least action principle.

By the use of Eq. (\ref{xyzG2}), 
it is shown that the Poisson bracket between $G(q,p)=\tilde{G}(x,y,z)$
and an arbitrary function $u(q,p)=\tilde{u}(x, y, z)$ vanishes such that
\begin{eqnarray}
\{G, u\}_{\mbox{\tiny{PB}}} 
&=& \frac{1}{2}
\sum_{i, j=1}^{3} \frac{\partial (\tilde{G}, \tilde{u})}
{\partial (x_i, x_j)} \{x_i, x_j\}_{\mbox{\tiny{PB}}}
= \frac{1}{2} \sum_{i, j, k=1}^{3} \varepsilon_{ijk} 
\frac{\partial (\tilde{G}, \tilde{u})}{\partial (x_i, x_j)} 
\frac{\partial \tilde{G}}{\partial x_k}\nonumber \\
&=& \frac{\partial (\tilde{G}, \tilde{u}, \tilde{G})}{\partial (x, y, z)} = 0~.
\label{Gu=0}
\end{eqnarray}
This means that $G$ is a constant.
We can eliminate the constant by redefining $G$, 
and the resulting $\tilde{G}(x,y,z)=0$ can be regarded 
as a $\it constraint$, which is induced by enlarging the phase space
from $(q, p)$ to $(x, y, z)$. 

Here we give two comments on the induced constraint $\tilde{G}(x,y,z)=0$.
First, in the case in which $\partial \tilde{G}/\partial z \ne 0$, 
we can solve $\tilde{G}(x,y,z)=0$ for $z$ and obtain $z=z(x,y)$.
Because the condition 
$\partial \tilde{G}/\partial z= \{x, y\}_{\mbox{\tiny{PB}}} \ne 0$
also enables us to express $q$ and $p$ as functions of $x$ and $y$,
the expression $z=z(x,y)$ can also be obtained  
by inserting $q=q(x,y)$ and $p=p(x,y)$ into $z=z(q,p)$.
Therefore the implicit form of the constraint $\tilde{G}(x,y,z)=0$
has an equivalent explicit form $z=z(x,y)$, which clearly shows that 
$z$ is a redundant variable in this case.
Second, $\tilde{H}(x,y,z)$ is not uniquely determined 
as a function of $x$, $y$, and $z$, i.e.,
we can add a term $\tilde{\lambda}(x,y,z)\tilde{G}(x,y,z)$ to $\tilde{H}(x,y,z)$,
where $\tilde{\lambda}(x,y,z)$ is some function.
If a Hamiltonian $\tilde{H}(x,y,z)$ satisfies 
$\tilde{H}(x,y,z)=H(q,p)$ and Eq. (\ref{H-eqf(N=3)}),
another Hamiltonian $\tilde{H}(x,y,z)+\tilde{\lambda}(x,y,z)\tilde{G}(x,y,z)$
also satisfies them.  
This is because the additional term $\tilde{\lambda}(x,y,z)\tilde{G}(x,y,z)$
always vanishes on the Nambu bracket.

It is straightforward to extend the above formulation
to the case with general $N(\ge 3)$.
We assume that at least $N-1$ of 
$\{x_i, x_j\}_{\mbox{\tiny{PB}}}$ $(i, j=1, \cdots, N)$ 
do not vanish.
In this case, the equation for any function 
$\tilde{f}(x_1, \cdots, x_N) = f(q, p)$
is written as 
\begin{eqnarray}
\frac{d \tilde{f}}{dt} = \frac{\partial(f, H)}{\partial(q, p)} 
= \frac{1}{2} \sum_{i, j=1}^{N} 
\frac{\partial (\tilde{f}, \tilde{H})}{\partial (x_{i}, x_{j})}
\{x_{i}, x_{j}\}_{\mbox{\tiny{PB}}}~,
\label{H-eq(gN)}
\end{eqnarray}
where $\tilde{H}(x_1, \cdots, x_N) = H(q, p)$.

Introducing functions 
$\tilde{G}_b=\tilde{G}_b(x_1, \cdots, x_N)$ $(b=1, \cdots, N-2)$
that satisfy the conditions
\begin{eqnarray}
\frac{1}{(N-2)!} \sum_{i_3 \cdots i_{N}=1}^{N}
\varepsilon_{i_1 i_2 i_3 \cdots i_{N}}
\frac{\partial (\tilde{G}_1, \cdots, \tilde{G}_{N-2})}
{\partial (x_{i_3}, \cdots, x_{i_{N}})}
= \{x_{i_1}, x_{i_2}\}_{\mbox{\tiny{PB}}}~,
\label{xiGb}
\end{eqnarray}
Eq. (\ref{H-eq(gN)}) is rewritten as the Nambu equation 
in the form of Eq. (\ref{N-eqf-N}),
\begin{eqnarray}
\frac{d \tilde{f}}{dt} = 
\{\tilde{f}, \tilde{H}, \tilde{G}_1, \cdots, \tilde{G}_{N-2}\}_{\mbox{\tiny{NB}}}~,
\label{H-eqf(gN)}
\end{eqnarray}
where we use the formula concerning Jacobians,
\begin{eqnarray}
\frac{\partial (\tilde{A}_1, \tilde{A}_2, \cdots, \tilde{A}_N)}
{\partial (x_1, x_2, \cdots, x_N)}
= \frac{1}{2(N-2)!} \sum_{i_1, i_2, i_3, \cdots i_{N}=1}^{N}
\varepsilon_{i_1 i_2 i_3 \cdots i_{N}}
\frac{\partial (\tilde{A}_1, \tilde{A}_2)}{\partial (x_{i_1}, x_{i_2})}
\frac{\partial (\tilde{A}_3, \cdots, \tilde{A}_{N})}
{\partial (x_{i_3}, \cdots, x_{i_{N}})}~.
\label{J(gN)}
\end{eqnarray}

By the use of Eq. (\ref{xiGb}), 
it is shown that the Poisson bracket between any of $N-2$ functions
$G_b(q,p)=\tilde{G}_b(x_1, x_2, \cdots, x_{N})$
and an arbitrary function $u(q,p)=\tilde{u}(x_1, x_2, \cdots, x_{N})$ 
vanishes such that
\begin{eqnarray}
\{G_b, u\}_{\mbox{\tiny{PB}}} 
&=& \frac{1}{2} \sum_{i_1, i_2=1}^{N} 
\frac{\partial (\tilde{G}_b, \tilde{u})}{\partial (x_{i_1}, x_{i_2})}
\{x_{i_1}, x_{i_2}\}_{\mbox{\tiny{PB}}}
\nonumber \\
&=& \frac{1}{2(N-2)!} \sum_{i_1, i_2, i_3, \cdots i_{N}=1}^{N} 
\varepsilon_{i_1 i_2 i_3 \cdots i_{N}}
\frac{\partial (\tilde{G}_b, \tilde{u})}{\partial (x_{i_1}, x_{i_2})}
\frac{\partial (\tilde{G}_1, \cdots, \tilde{G}_{N-2})}
{\partial (x_{i_3}, \cdots, x_{i_{N}})}
\nonumber \\
&=& \frac{\partial (\tilde{G}_b, \tilde{u}, \tilde{G}_1, \cdots, \tilde{G}_{N-2})}
{\partial (x_{1}, x_{2}, x_{3}, \cdots, x_{N})} = 0~.
\label{Gbu=0}
\end{eqnarray}
Hence $G_b$ are constants.
We can eliminate the constants by redefining $G_b$, 
and the resulting $\tilde{G}_b(x_1,x_2,\cdots, x_{N})=0$ can be regarded 
as {\it induced constraints}, which are associated with enlarging the phase space
from $(q, p)$ to $(x_1,x_2,\cdots, x_{N})$. 

In this way, 
{\it Hamiltonian systems can be formulated as Nambu systems 
by the use of $N$ variables $x_i=x_i(q, p)$ $(i=1, 2, \cdots, N)$.
The variables form an $N$-plet, and  
the $N-1$ Hamiltonians are given by the original Hamiltonian
$\tilde{H}(x_1,x_2,\cdots, x_{N})=H(q,p)$ 
and induced constraints $\tilde{G}_b(x_1,x_2,\cdots, x_{N})=0$
$(b=1, \cdots, N-2)$.}
Note that $\tilde{H}(x_1,x_2,\cdots, x_{N})$ is not uniquely determined, 
because of the freedom to add a term 
$\sum_b\tilde{\lambda}_b(x_1,x_2,\cdots, x_{N})\tilde{G}_b(x_1,x_2,\cdots, x_{N})$
to $\tilde{H}(x_1,x_2,\cdots, x_{N})$.
Here $\tilde{\lambda}_b(x_1,x_2,\cdots, x_{N})$ 
are some functions.

\subsubsection{Examples}
\label{Examples}

Here we present two simple examples to show how induced constraints
are obtained for given multiplets.\\
(a) $N=3$\\
Consider composite variables,
\begin{eqnarray}
x=\frac{1}{4}\left(q^2 - p^2\right)~,~~ 
y=\frac{1}{4}\left(q^2 + p^2\right)~,~~ z=\frac{1}{2}qp~,
\label{Ex1}
\end{eqnarray}
which satisfy the following relations:
\begin{eqnarray}
\{x, y\}_{\mbox{\tiny{PB}}} = z~,~~
 \{y, z\}_{\mbox{\tiny{PB}}} = x~,~~ \{z, x\}_{\mbox{\tiny{PB}}} = -y~.
\label{Ex1-PB}
\end{eqnarray}
Then the conditions (Eq. (\ref{xyzG})) become
\begin{eqnarray}
\frac{\partial \tilde{G}}{\partial x} = x~,~~ 
\frac{\partial \tilde{G}}{\partial y} = -y~,~~ 
\frac{\partial \tilde{G}}{\partial z} = z~,
\label{Ex1-dG}
\end{eqnarray} 
and $\tilde{G}$ is obtained by
\begin{eqnarray}
\tilde{G}= \frac{1}{2}\left(x^2 - y^2 + z^2\right)+C~,
\label{Ex1-G}
\end{eqnarray}
where $C$ is a constant.
Redefining $\tilde{G}$ as $\tilde{G}-C$,
we obtain the induced constraint $\tilde{G}(x,y,z)=G(q,p)=0$.

~~\\
(b) $N=4$\\
Consider variables including composite ones,
\begin{eqnarray}
x_1=q~,~~ x_2=p~,~~ x_3=x_3(q, p)~,~~ x_4=x_4(q, p)~,
\label{Ex2}
\end{eqnarray}
which satisfy the following relations:
\begin{eqnarray}
&& \{x_1, x_2\}_{\mbox{\tiny{PB}}} 
= 1~,~~ \{x_1, x_3\}_{\mbox{\tiny{PB}}} = \frac{\partial x_3}{\partial p}~,~~ 
\{x_1, x_4\}_{\mbox{\tiny{PB}}} = \frac{\partial x_4}{\partial p}~,~~
\nonumber \\
&& \{x_2, x_3\}_{\mbox{\tiny{PB}}} = 
-\frac{\partial x_3}{\partial q}~,~~ 
\{x_2, x_4\}_{\mbox{\tiny{PB}}} = -\frac{\partial x_4}{\partial q}~,~~ 
\nonumber \\
&&\{x_3, x_4\}_{\mbox{\tiny{PB}}} = 
\frac{\partial x_3}{\partial q} \frac{\partial x_4}{\partial p}
- \frac{\partial x_3}{\partial p} \frac{\partial x_4}{\partial q}~.
\label{Ex2-PB}
\end{eqnarray}
Then the conditions (Eq. (\ref{xiGb})) become
\begin{eqnarray}
\sum_{i_3, i_4=1}^{4} \varepsilon_{i_1 i_2 i_3 i_4}
\frac{\partial \tilde{G}_1}{\partial x_{i_3}} 
\frac{\partial \tilde{G}_2}{\partial x_{i_4}}
= \{x_{i_1}, x_{i_2}\}_{\mbox{\tiny{PB}}}~,
\label{Ex2-dG}
\end{eqnarray} 
and $\tilde{G}_1$ and $\tilde{G}_2$ are given by
\begin{eqnarray}
\tilde{G}_1= x_3-x_3(x_1, x_2)+C_1~,~~ \tilde{G}_2= x_4-x_4(x_1, x_2)+C_2~.
\label{Ex2-G}
\end{eqnarray}
where $C_1$ and $C_2$ are constants.
By redefining $G_1$ and $G_2$ to eliminate the constants,
we obtain the induced constraints 
$\tilde{G}_1(x_1,x_2,x_3,x_4)=G_1(q,p)=0$ and 
$\tilde{G}_2(x_1,x_2,x_3,x_4)=G_2(q,p)=0$.

\subsection{Many degrees of freedom systems}
\label{Many degrees of freedom systems}
Let us extend our formulation
to Hamiltonian systems with many degrees of freedom.
Consider a Hamiltonian system described by 
$n$ sets of canonical doublets 
$(q_{\mbox{\tiny(k)}}, p_{\mbox{\tiny(k)}})$ (k$=1, 2, \cdots, n$).
As is the case with $n=1$ given in Sect. 2.2,
hidden Nambu structure can also be found in this system. 
Here we present the $N=3$ case, i.e.,  the case with $n$ sets of triplets 
$x_{i{\mbox{\tiny(k)}}} = x_{i{\mbox{\tiny(k)}}}
(q_{\mbox{\tiny(k)}}, p_{\mbox{\tiny(k)}})$ $(i=1,2,3)$.
Generalization to the $N(\ge 3)$ cases is straightforward.

\subsubsection{Dynamics}
\label{Dynamics}

In this system, the Poisson bracket of $A$ and $B$ is defined as
\begin{eqnarray}
\{A, B\}_{\mbox{\tiny{PB}}} 
\equiv \sum_{{\rm k}=1}^{n} 
\left(\frac{\partial A}{\partial q_{\mbox{\tiny(k)}}}
\frac{\partial B}{\partial p_{\mbox{\tiny(k)}}}
 - \frac{\partial A}{\partial p_{\mbox{\tiny(k)}}}
 \frac{\partial B}{\partial q_{\mbox{\tiny(k)}}}\right)~,
\label{PB-n}
\end{eqnarray}
and the Hamilton's equation of motion for any function 
$f=f(q_{\mbox{\tiny(1)}}, p_{\mbox{\tiny(1)}}, 
\cdots, q_{\mbox{\tiny({\it n})}}, p_{\mbox{\tiny({\it n})}})$ 
can be written as
\begin{eqnarray}
\frac{d f}{dt} = \{f, H\}_{\mbox{\tiny{PB}}}~,
\label{H-eqf-n}
\end{eqnarray}
where $H=H(q_{\mbox{\tiny(1)}}, p_{\mbox{\tiny(1)}}, 
\cdots, q_{\mbox{\tiny({\it n})}}, p_{\mbox{\tiny({\it n})}})$
is the Hamiltonian of the system.
On the other hand, the Nambu bracket of $\tilde{A}$, $\tilde{B}$, 
and $\tilde{C}$ is defined as
\begin{eqnarray}
\{\tilde{A}, \tilde{B}, \tilde{C}\}_{\mbox{\tiny{NB}}} 
\equiv \sum_{{\rm k}=1}^{n} 
\frac{\partial (\tilde{A}, \tilde{B}, \tilde{C})}
{\partial (x_{\mbox{\tiny(k)}}, y_{\mbox{\tiny(k)}}, z_{\mbox{\tiny(k)}})}~,
\label{NB-n}
\end{eqnarray}
where 
$x_{\mbox{\tiny(k)}} = x_{1\mbox{\tiny(k)}}$, 
$y_{\mbox{\tiny(k)}} = x_{2\mbox{\tiny(k)}}$, and 
$z_{\mbox{\tiny(k)}} = x_{3\mbox{\tiny(k)}}$. 
Then the Nambu equation for any function 
$\tilde{f}=
\tilde{f}(x_{\mbox{\tiny(1)}}, y_{\mbox{\tiny(1)}}, \cdots, 
z_{\mbox{\tiny({\it n})}})$ can be written as
\begin{eqnarray}
\frac{d \tilde{f}}{dt} = \{\tilde{f}, \tilde{H}, \tilde{G}\}_{\mbox{\tiny{NB}}}~.
\label{N-eqf-n}
\end{eqnarray}
Here 
$\tilde{H}=
\tilde{H}(x_{\mbox{\tiny(1)}}, y_{\mbox{\tiny(1)}}, 
\cdots, z_{\mbox{\tiny({\it n})}})
=H(q_{\mbox{\tiny(1)}}, p_{\mbox{\tiny(1)}}, \cdots, 
q_{\mbox{\tiny({\it n})}}, p_{\mbox{\tiny({\it n})}})$ 
is the Hamiltonian and 
$\tilde{G}=\tilde{G}(x_{\mbox{\tiny(1)}}, y_{\mbox{\tiny(1)}}, 
\cdots, z_{\mbox{\tiny({\it n})}})
= \sum_{\rm k} \tilde{G}_{\mbox{\tiny(k)}}
(x_{\mbox{\tiny(k)}}, y_{\mbox{\tiny(k)}}, z_{\mbox{\tiny(k)}})$
is the sum of the induced constraints 
that satisfy the conditions
\begin{eqnarray}
\frac{\partial \tilde{G}_{\mbox{\tiny(k)}}}{\partial x_{\mbox{\tiny(k)}}} = 
\frac{\partial(y_{\mbox{\tiny(k)}}, z_{\mbox{\tiny(k)}})}
{\partial(q_{\mbox{\tiny(k)}}, p_{\mbox{\tiny(k)}})}~,~~
\frac{\partial \tilde{G}_{\mbox{\tiny(k)}}}{\partial y_{\mbox{\tiny(k)}}} = 
\frac{\partial(z_{\mbox{\tiny(k)}}, x_{\mbox{\tiny(k)}})}
{\partial(q_{\mbox{\tiny(k)}}, p_{\mbox{\tiny(k)}})}~,~~
\frac{\partial \tilde{G}_{\mbox{\tiny(k)}}}{\partial z_{\mbox{\tiny(k)}}} = 
\frac{\partial(x_{\mbox{\tiny(k)}}, y_{\mbox{\tiny(k)}})}
{\partial(q_{\mbox{\tiny(k)}}, p_{\mbox{\tiny(k)}})}~.
\label{xyzG-n}
\end{eqnarray}
Note that the induced constraints are defined so as to be zero, 
$\tilde{G}_{\mbox{\tiny(k)}}
(x_{\mbox{\tiny(k)}}, y_{\mbox{\tiny(k)}}, z_{\mbox{\tiny(k)}})
=G_{\mbox{\tiny(k)}}(q_{\mbox{\tiny(k)}}, p_{\mbox{\tiny(k)}})=0$, 
and the Hamiltonian is not uniquely determined 
because of the freedom to add a linear combination of 
$\tilde{G}_{\mbox{\tiny(k)}}$ to $\tilde{H}$.

The $3n$ variables $x_{i\mbox{\tiny(k)}}$ satisfy the relations
\begin{eqnarray}
\{
x_{i_1{\mbox{\tiny($\rm k_1$)}}}, 
x_{i_2{\mbox{\tiny($\rm k_2$)}}}, 
x_{i_3{\mbox{\tiny($\rm k_3$)}}}
\}_{\mbox{\tiny{NB}}} 
&=& \varepsilon_{i_1 i_2 i_3}~~~~\mbox{for ~$\rm k_1=k_2=k_3$}~,
\label{NB-n-rel1}\\
\{
x_{i_1{\mbox{\tiny($\rm k_1$)}}}, 
x_{i_2{\mbox{\tiny($\rm k_2$)}}}, 
x_{i_3{\mbox{\tiny($\rm k_3$)}}}
\}_{\mbox{\tiny{NB}}} 
&=& 0~~~~~~~~~~\mbox{otherwise}.
\label{NB-n-rel2}
\end{eqnarray}
The first type of relation (Eq. (\ref{NB-n-rel1})) is invariant 
under the time evolution (Eq. (\ref{N-eqf-n}))
irrespective of the form of $\tilde{H}$.
To be more specific,
for infinitesimal transformations
$x_{i\mbox{\tiny(k)}} \to x'_{i\mbox{\tiny(k)}}=
x_{i\mbox{\tiny(k)}}+(dx_{i\mbox{\tiny(k)}}/dt) dt$,
\begin{eqnarray}
\{x'_{\mbox{\tiny(k)}}, y'_{\mbox{\tiny(k)}}, z'_{\mbox{\tiny(k)}}\}
_{\mbox{\tiny{NB}}} 
= 1~\label{NB-n-rel'}
\end{eqnarray}
hold. We can also show an important relation,
\begin{eqnarray}
\frac{\partial(x'_{\mbox{\tiny(1)}}, y'_{\mbox{\tiny(1)}}, z'_{\mbox{\tiny(1)}}, 
\cdots, 
x'_{\mbox{\tiny({\it n})}}, y'_{\mbox{\tiny({\it n})}}, z'_{\mbox{\tiny({\it n})}})}
{\partial(x_{\mbox{\tiny(1)}}, y_{\mbox{\tiny(1)}}, z_{\mbox{\tiny(1)}}, \cdots, 
x_{\mbox{\tiny({\it n})}}, y_{\mbox{\tiny({\it n})}}, z_{\mbox{\tiny({\it n})}})}
= 1~,\label{LT}
\end{eqnarray}
which guarantees the Liouville theorem, 
the conservation law of the phase space volume 
under time development.
On the other hand, the second type of relation (Eq. (\ref{NB-n-rel2})) 
does not always hold,
unless there is no interaction between the $n$ subsystems,
i.e.,
$\tilde{H}$ has a form such as
$\tilde{H}(x_{\mbox{\tiny(1)}}, y_{\mbox{\tiny(1)}}, 
\cdots, z_{\mbox{\tiny({\it n})}})
= \sum_{\rm k} \tilde{H}_{\mbox{\tiny(k)}}
(x_{\mbox{\tiny(k)}}, y_{\mbox{\tiny(k)}}, z_{\mbox{\tiny(k)}})$.

\subsubsection{Partition functions}
\label{Partition functions}

It is well known that 
the partition function $Z_{\rm H}$ for a canonical ensemble
of the Hamiltonian system 
$(q_{\mbox{\tiny(1)}}, p_{\mbox{\tiny(1)}}, \cdots, 
q_{\mbox{\tiny({\it n})}}, p_{\mbox{\tiny({\it n})}})$
is defined as
\begin{eqnarray}
Z_{\rm H} \equiv 
\iint\!\!\cdot\!\cdot\!\cdot\!\!\int 
\prod_{{\rm k}=1}^{n} dq_{\mbox{\tiny(k)}} dp_{\mbox{\tiny(k)}} e^{-\beta H}~,
\label{ZH}
\end{eqnarray}
where $\beta = 1/(k_BT)$ is the inverse temperature
made up of the Boltzmann constant $k_B$ and the temperature $T$.
Here we study the partition function $Z_{\rm N}$ 
for an ensemble of the Nambu system
$(x_{\mbox{\tiny(1)}}, y_{\mbox{\tiny(1)}}, \cdots, z_{\mbox{\tiny({\it n})}})$
hidden in the Hamiltonian system.

First let us conjecture
the form of $Z_{\rm N}$ on physical grounds. 
Since $\tilde{H}=H$, $Z_{\rm N}$ must contain 
the $\lq\lq$Boltzmann weight" such as $e^{-\beta \tilde{H}}$.
The other Hamiltonian $\tilde{G}$ is the sum of the constraints 
$\tilde{G}_{\mbox{\tiny(k)}} 
(x_{\mbox{\tiny(k)}}, y_{\mbox{\tiny(k)}}, z_{\mbox{\tiny(k)}})
= G_{\mbox{\tiny(k)}} (q_{\mbox{\tiny(k)}}, p_{\mbox{\tiny(k)}})= 0$,
and therefore there should be 
delta functions such as $\delta(\tilde{G}_{\mbox{\tiny(k)}})$ 
in $Z_{\rm N}$.
Furthermore, $Z_{\rm N}$ must contain the volume element 
$\prod_{{\rm k}=1}^{n}
dx_{\mbox{\tiny(k)}} dy_{\mbox{\tiny(k)}} dz_{\mbox{\tiny(k)}}$
from the Liouville theorem.

On the basis of the above observations, 
it is expected that $Z_{\rm N}$ should have a form such that
\begin{eqnarray}
Z_{\rm N} &\equiv& 
\iint\!\!\cdot\!\cdot\!\cdot\!\!\int 
\prod_{{\rm k}=1}^{n}
dx_{\mbox{\tiny(k)}} dy_{\mbox{\tiny(k)}} dz_{\mbox{\tiny(k)}} 
\delta(\tilde{G}_{\mbox{\tiny(k)}}) e^{-\beta \tilde{H}}
\label{ZN1}\\
&=& \iint\!\!\cdot\!\cdot\!\cdot\!\!\int  
\prod_{{\rm k}=1}^{n} 
dx_{\mbox{\tiny(k)}} dy_{\mbox{\tiny(k)}} dz_{\mbox{\tiny(k)}}  
\int_{-\infty}^{\infty} \frac{d\gamma_{\mbox{\tiny(k)}}}{2\pi} 
e^{-\beta \tilde{H}
-i \sum_{\rm k} \gamma_{\mbox{\tiny(k)}} \tilde{G}_{\mbox{\tiny(k)}}}~.
\label{ZN2}
\end{eqnarray}
We can derive $Z_{\rm H}$ (Eq. (\ref{ZH})) from this expression for $Z_{\rm N}$.
For example, let us consider the case that
$\partial \tilde{G}_{\mbox{\tiny(k)}}/\partial z_{\mbox{\tiny(k)}} \ne 0$.
We assume that there are 
$N_{\mbox{\tiny k}}$ solutions of $\tilde{G}_{\mbox{\tiny(k)}}=0$,
$z_{\mbox{\tiny(k)}}^{(a_{\mbox{\tiny k}})}$ 
$(a_{\mbox{\tiny k}} = 1, 2, \cdots, N_{\mbox{\tiny k}})$,
and all of them satisfy the conditions (Eq. (\ref{xyzG-n})).
Then using the formula for the delta function and the change of variables,
Eq. (\ref{ZN1}) becomes
\begin{eqnarray}
Z_{\rm N} &=& 
\iint\!\!\cdot\!\cdot\!\cdot\!\!\int 
\prod_{{\rm k}=1}^{n} 
dx_{\mbox{\tiny(k)}} dy_{\mbox{\tiny(k)}} dz_{\mbox{\tiny(k)}} 
\sum_{a_{\mbox{\tiny k}}=1}^{N_{\mbox{\tiny k}}}
\delta(z_{\mbox{\tiny(k)}}-
z_{\mbox{\tiny(k)}}^{(a_{\mbox{\tiny k}})}(x_{\mbox{\tiny(k)}},y_{\mbox{\tiny(k)}})) 
\left|\frac{\partial \tilde{G}_{\mbox{\tiny(k)}}}
{\partial z_{\mbox{\tiny(k)}}}\right|^{-1} e^{-\beta \tilde{H}}
\nonumber \\
&=& \iint\!\!\cdot\!\cdot\!\cdot\!\!\int
\prod_{{\rm k}=1}^{n} 
dx_{\mbox{\tiny(k)}} dy_{\mbox{\tiny(k)}} dz_{\mbox{\tiny(k)}} 
\sum_{a_{\mbox{\tiny k}}=1}^{N_{\mbox{\tiny k}}}
\delta(z_{\mbox{\tiny(k)}}-
z_{\mbox{\tiny(k)}}^{(a_{\mbox{\tiny k}})}(x_{\mbox{\tiny(k)}},y_{\mbox{\tiny(k)}})) 
\left|\frac{\partial (x_{\mbox{\tiny(k)}}, y_{\mbox{\tiny(k)}})}
{\partial (q_{\mbox{\tiny(k)}}, p_{\mbox{\tiny(k)}})}\right|^{-1}
e^{-\beta \tilde{H}}
\nonumber \\
&=& {\mathcal N}
\iint\!\!\cdot\!\cdot\!\cdot\!\!\int
\prod_{{\rm k}=1}^{n} dq_{\mbox{\tiny(k)}} dp_{\mbox{\tiny(k)}} e^{-\beta H} 
= {\mathcal N} Z_{\rm H}~,
\label{ZH=ZN}
\end{eqnarray}
where ${\mathcal N}=\prod_{{\rm k}=1}^{n}N_{\mbox{\tiny k}}$
is a constant normalization factor.
This factor is irrelevant to the evaluation of physical quantities.

It is natural to require that $Z_{\rm N}$ should agree with $Z_{\rm H}$
(up to some normalization factor),
because we just describe the same physical system 
using different formulations.
It should be noted here that both expressions for $Z_{\rm N}$ 
 (Eq. (\ref{ZN1}) or Eq. (\ref{ZN2})) 
are different from that proposed in Ref. \cite{N}.
This comes from the fact that the Nambu mechanics considered here
is an effective one induced by the redundancy of the variables.


Finally, we just give the result for the case of general $N(\ge 3)$.
The possible form of the partition function is given by 
\begin{eqnarray}
Z_{\rm N} = \iint\!\!\cdot\!\cdot\!\cdot\!\!\int
\prod_{{\rm k}=1}^{n} 
dx_{1\mbox{\tiny{(k)}}} dx_{2\mbox{\tiny{(k)}}} 
\cdots dx_{N\mbox{\tiny{(k)}}} 
\delta(\tilde{G}_{1\mbox{\tiny{(k)}}}) \delta(\tilde{G}_{2\mbox{\tiny{(k)}}}) 
\cdots \delta(\tilde{G}_{N-2\mbox{\tiny{(k)}}}) e^{-\beta \tilde{H}}~,
\label{ZN-N}
\end{eqnarray}
where $\tilde{G}_{b\mbox{\tiny{(k)}}}=0$ $(b = 1, 2, \cdots, N-2)$ 
are induced constraints.
This expression should agree with $Z_{\rm H}$ (Eq. (\ref{ZH})) 
up to some constant normalization factor.

\subsection{Generalized Nambu equations}
\label{Generalized Nambu equations}

We generalize our formulation to include a specific case
that all multiplets share some variables. 
In such a case, a generalization of the Nambu equation would be required. 

Let us describe a Hamiltonian system with $n$ sets of 
canonical doublets $(q_{\mbox{\tiny(k)}}, p_{\mbox{\tiny(k)}})$ 
$({\rm k}=1, \cdots, n)$
using $2n+m$ variables $w_{\ell}$ $(\ell=1, \cdots, 2n+m)$.
We classify the variables $w_{\ell}$ into two groups,
$x_{a}$ $(a=1, \cdots, 2n)$ and $z_s$ $(s=1, \cdots, m)$,
where $x_a$ are assumed to 
satisfy $\det \{x_a, x_b\}_{\mbox{\tiny{PB}}} \ne 0$.
Note that the classification of variables is not unique.

First we study the case with $m=0$ for completeness.
In this case, the equation for any function 
$\tilde{f}(x_1, \cdots, x_{2n}) = 
f(q_{\mbox{\tiny(1)}}, p_{\mbox{\tiny(1)}}, 
\cdots, q_{\mbox{\tiny({\it n})}}, p_{\mbox{\tiny({\it n})}})$
can be written as
\begin{eqnarray}
\frac{d \tilde{f}}{dt} = \sum_{{\rm k}=1}^{n} 
\frac{\partial(f, H)}{\partial(q_{\mbox{\tiny(k)}}, p_{\mbox{\tiny(k)}})}
= \frac{1}{2} \sum_{{\rm k}=1}^{n} 
\sum_{{a, b}=1}^{2n} \frac{\partial(\tilde{f}, \tilde{H})}{\partial(x_a, x_b)}
\frac{\partial(x_a, x_b)}{\partial(q_{\mbox{\tiny(k)}}, p_{\mbox{\tiny(k)}})} 
= \sum_{{a, b}=1}^{2n} \tilde{g}_{ab} 
\frac{\partial(\tilde{f}, \tilde{H})}{\partial(x_a, x_b)}~,
\label{H-eq(m=0)}
\end{eqnarray}
where $\tilde{H}=\tilde{H}(x_1, \cdots, x_{2n}) = 
H(q_{\mbox{\tiny(1)}}, p_{\mbox{\tiny(1)}}, 
\cdots, q_{\mbox{\tiny({\it n})}}, p_{\mbox{\tiny({\it n})}})$
and $\tilde{g}_{ab}$ is defined as
\begin{eqnarray}
\tilde{g}_{ab}(x_1, \cdots, x_{2n}) 
= g_{ab}(q_{\mbox{\tiny(1)}}, p_{\mbox{\tiny(1)}}, 
\cdots, q_{\mbox{\tiny({\it n})}}, p_{\mbox{\tiny({\it n})}})
\equiv \frac{1}{2}\sum_{{\rm k}=1}^{n} 
\frac{\partial(x_a, x_b)}{\partial(q_{\mbox{\tiny(k)}}, p_{\mbox{\tiny(k)}})}
= \frac{1}{2}\{x_a, x_b\}_{\mbox{\tiny{PB}}}~.
\label{Gab}
\end{eqnarray}
The $\tilde{g}_{ab}$ plays the role of a metric tensor,
because it transforms under a change of variables $x_a \to x'_a$ as follows:
\begin{eqnarray}
\tilde{g}'_{ab}(x'_1, \cdots, x'_{2n}) 
= \sum_{{c, d}=1}^{2n} \frac{\partial x'_a}{\partial x_c} 
\frac{\partial x'_b}{\partial x_d} ~\tilde{g}_{cd}(x_1, \cdots, x_{2n})~.
\label{G'ab}
\end{eqnarray}
In the case in which $\tilde{g}_{ab}$ depends on $x_a$,
neither the transformation 
$(q_{\mbox{\tiny(1)}}, p_{\mbox{\tiny(1)}}, 
\cdots, q_{\mbox{\tiny({\it n})}}, p_{\mbox{\tiny({\it n})}})
\to (x_1, \cdots, x_{2n})$
nor the time evolution of $x_a$ is a canonical transformation.
The latter means that the Liouville theorem in general 
does not hold for the dynamics of $x_a$.
This fact reminds us of the superiority of canonical variables.

Now let us proceed to the case with $m \ge 1$.
The equation for a function
$\tilde{f}(w_1, \cdots, w_{2n+m}) = 
f(q_{\mbox{\tiny(1)}}, p_{\mbox{\tiny(1)}}, 
\cdots, q_{\mbox{\tiny({\it n})}}, p_{\mbox{\tiny({\it n})}})$
can be written as
\begin{eqnarray}
\frac{d \tilde{f}}{dt} 
&=& \frac{1}{2} \sum_{{a, b}=1}^{2n} \frac{\partial(\tilde{f}, \tilde{H})}
{\partial(x_a, x_b)}\{x_a, x_b\}_{\mbox{\tiny{PB}}} \nonumber \\
&&+ \sum_{a=1}^{2n}\sum_{s=1}^{m} \frac{\partial(\tilde{f}, \tilde{H})}
{\partial(x_a, z_s)}\{x_a, z_s\}_{\mbox{\tiny{PB}}}
+ \frac{1}{2} \sum_{s, t=1}^{m} \frac{\partial(\tilde{f}, \tilde{H})}
{\partial(z_s, z_t)}\{z_s, z_t\}_{\mbox{\tiny{PB}}}~,
\label{H-eq(m)}
\end{eqnarray}
where $\tilde{H}= \tilde{H}(w_1, \cdots, w_{2n+m}) 
=H(q_{\mbox{\tiny(1)}}, p_{\mbox{\tiny(1)}}, 
\cdots, q_{\mbox{\tiny({\it n})}}, p_{\mbox{\tiny({\it n})}})$.
Introducing functions $\tilde{G}_s$ $(s=1, \cdots, m)$
and $\tilde{g}^{(m)}_{ab}$
that satisfy the following relations,
\begin{eqnarray}
\frac{1}{2} \{x_a, x_b\}_{\mbox{\tiny{PB}}}
&=& 
\tilde{g}^{(m)}_{ab} \frac{\partial (\tilde{G}_1, \cdots, \tilde{G}_{m})}
{\partial (z_{1}, \cdots, z_{m})}~,~~
\label{phi-2n+m1}\\
\frac{1}{2}\{x_a, z_{s}\}_{\mbox{\tiny{PB}}}
&=& 
-\sum_{b=1}^{2n} \tilde{g}^{(m)}_{ab} 
\frac{\partial (\tilde{G}_1, \cdots, \tilde{G}_{s-1},
\tilde{G}_{s}, \tilde{G}_{s+1}, \cdots, \tilde{G}_{m})}
{\partial (z_{1}, \cdots, z_{s-1}, x_b, z_{s+1}, \cdots,z_{m})}~,~~
\label{phi-2n+m2}\\
\{z_s, z_t\}_{\mbox{\tiny{PB}}}
&=&\!\!\!\!
\sum_{a,b=1}^{2n}\tilde{g}^{(m)}_{ab} 
\frac{\partial (\tilde{G}_1, \cdots, 
\tilde{G}_{s-1}, \tilde{G}_{s}, \tilde{G}_{s+1}, \cdots, 
\tilde{G}_{t-1}, \tilde{G}_{t}, \tilde{G}_{t+1}, \cdots, \tilde{G}_{m})}
{\partial (z_{1}, \cdots, z_{s-1}, x_{a}, z_{s+1},
\cdots, z_{t-1}, x_{b}, z_{t+1}, \cdots, z_{m})}~,
\label{phi-2n+m3}
\end{eqnarray}
where $s<t$, Eq. (\ref{H-eq(m)}) can be rewritten as
\begin{eqnarray}
\frac{d \tilde{f}}{dt}
= \sum_{a, b=1}^{2n} \tilde{g}^{(m)}_{ab} 
\frac{\partial (\tilde{f}, \tilde{H}, \tilde{G}_1, \cdots, \tilde{G}_{m})}
{\partial (x_a, x_b, z_1, \cdots, z_m)}
\label{N-eqf(m)}
\end{eqnarray}
using a formula concerning Jacobians.

By the use of Eqs. (\ref{phi-2n+m1})--(\ref{phi-2n+m3}), 
it is shown that the Poisson bracket between any of the $m$ functions
$G_s(q_{\mbox{\tiny(1)}}, p_{\mbox{\tiny(1)}}, 
\cdots, q_{\mbox{\tiny({\it n})}}, p_{\mbox{\tiny({\it n})}})
=\tilde{G}_s(w_1, \cdots, w_{2n+m})$
and an arbitrary function 
$u(q_{\mbox{\tiny(1)}}, p_{\mbox{\tiny(1)}}, 
\cdots, q_{\mbox{\tiny({\it n})}}, p_{\mbox{\tiny({\it n})}})
=\tilde{u}(w_1, \cdots, w_{2n+m})$ vanishes such that
\begin{eqnarray}
\{G_s, u\}_{\mbox{\tiny{PB}}} 
&=& \frac{1}{2} \sum_{a, b=1}^{2n} 
\frac{\partial(\tilde{G}_s, \tilde{u})}
{\partial(x_a, x_b)}\{x_a, x_b\}_{\mbox{\tiny{PB}}}
\nonumber \\
&& 
+ \sum_{a=1}^{2n} \sum_{s=1}^{m} 
\frac{\partial(\tilde{G}_s, \tilde{u})}
{\partial(x_a, z_s)}\{x_a, z_s\}_{\mbox{\tiny{PB}}}
+ \frac{1}{2} \sum_{s, t=1}^{m} 
\frac{\partial(\tilde{G}_s, \tilde{u})}
{\partial(z_s, z_t)}\{z_s, z_t\}_{\mbox{\tiny{PB}}}
\nonumber \\
&=& \sum_{a, b=1}^{2n} 
\tilde{g}^{(m)}_{ab} 
\frac{\partial (\tilde{G}_s, \tilde{u}, \tilde{G}_1, \cdots, \tilde{G}_{m})}
{\partial (x_a, x_b, z_1, \cdots, z_m)} = 0~.
\label{phiu=0}
\end{eqnarray}
Hence 
$G_s(q_{\mbox{\tiny(1)}}, p_{\mbox{\tiny(1)}}, 
\cdots, q_{\mbox{\tiny({\it n})}}, p_{\mbox{\tiny({\it n})}})$ 
are constants and, if necessary,
we can define $G_s=\tilde{G}_s=0$ by shifting constants. 
We refer to Eq. (\ref{N-eqf(m)}) as the generalized Nambu equation.
Note that the Liouville theorem does not hold in general
for the dynamics given by this equation.
This unfavorable property is a result of two factors:
Eq. (\ref{N-eqf(m)}) has $x_{a}$-dependent $\tilde{g}^{(m)}_{ab}$
and multiplets in Eq. (\ref{N-eqf(m)}) share common variables $z_{s}$.   
The latter means that it is difficult to define 
an appropriate phase space volume.

One of the non-vanishing components of $\tilde{g}^{(m)}_{ab}$ 
can be set to $\frac{1}{2}$ by redefinition of constraints $\tilde{G}_s$.
For example, in the case in which $n=1$,
we can set $\tilde{g}^{(m)}_{12}=\frac{1}{2}$ 
(and $\tilde{g}^{(m)}_{21}=-\frac{1}{2}$) by redefining $\tilde{G}_s$,
and Eq. (\ref{N-eqf(m)}) reduces to 
the Nambu equation (Eq. (\ref{H-eqf(gN)})) with $N=2+m$. 

Finally, we consider the case in which the variables $x_a$ and $z_s$ are
further classified into $M$ $\lq\lq$irreducible" sets,
$\{x_{a^1}^{\mbox{\tiny{({\rm 1})}}}, 
z_{s^1}^{\mbox{\tiny{({\rm 1})}}}\} 
\bigoplus 
\{x_{a^2}^{\mbox{\tiny{({\rm 2})}}}, 
z_{s^2}^{\mbox{\tiny{({\rm 2})}}}\} 
\bigoplus \cdots \bigoplus 
\{x_{a^M}^{\mbox{\tiny{({\it M})}}}, 
z_{s^M}^{\mbox{\tiny{({\it M})}}}\}$,
where $a^i=1,\cdots,2n^i$ ($\sum_{i=1}^{M}n^i=n$)
and $s^i=1,\cdots,m^i$ ($\sum_{i=1}^{M}m^i=m$).
Here $\lq\lq$irreducible" means that
the Poisson bracket between any two elements that belong to different sets
vanishes, i.e.,
$\{x_{a^i}^{\mbox{\tiny{({\it i})}}}, 
x_{a^j}^{\mbox{\tiny{({\it j})}}}\}_{\mbox{\tiny{PB}}}=0$, 
$\{x_{a^i}^{\mbox{\tiny{({\it i})}}}, 
z_{s^j}^{\mbox{\tiny{({\it j})}}}\}_{\mbox{\tiny{PB}}}=0$, 
and 
$\{z_{s^i}^{\mbox{\tiny{({\it i})}}}, 
z_{s^j}^{\mbox{\tiny{({\it j})}}}\}_{\mbox{\tiny{PB}}}=0$ 
for $i \ne j$.
Note that this classification is not unique, either.
The equation of motion 
for any function $\tilde{f}(w_1, \cdots, w_{2n+m})$ can be expressed 
in the form of the generalized Nambu equation,
\begin{eqnarray}
\frac{d \tilde{f}}{dt}
= \sum_{i=1}^{M}
\sum_{a^i, b^i=1}^{2n^i} \tilde{g}^{(m^i)}_{a^i b^i} 
\frac{\partial (\tilde{f}, \tilde{H}, 
\tilde{G}_{1}^{\mbox{\tiny{({\it i})}}}, 
\cdots, \tilde{G}_{m^i}^{\mbox{\tiny{({\it i})}}})}
{\partial (x_{a^i}^{\mbox{\tiny{({\it i})}}}, 
x_{b^i}^{\mbox{\tiny{({\it i})}}}, 
z_{1}^{\mbox{\tiny{({\it i})}}}, \cdots, 
z_{m^i}^{\mbox{\tiny{({\it i})}}})}~.
\label{N-eqf(m)2}
\end{eqnarray}
Here $\tilde{G}_{s^i}^{\mbox{\tiny{({\it i})}}}$ 
and $\tilde{g}^{(m^i)}_{a^ib^i}$ 
should satisfy the following conditions:
\begin{eqnarray}
\frac{1}{2} \{x_{a^i}^{\mbox{\tiny{({\it i})}}}, 
x_{b^i}^{\mbox{\tiny{({\it i})}}}\}_{\mbox{\tiny{PB}}}
&=& 
\tilde{g}^{(m^i)}_{a^ib^i} 
\frac{\partial (\tilde{G}_{1}^{\mbox{\tiny{({\it i})}}}, 
\cdots, \tilde{G}_{m^i}^{\mbox{\tiny{({\it i})}}})}
{\partial (z_{1}^{\mbox{\tiny{({\it i})}}}, \cdots, 
z_{m^i}^{\mbox{\tiny{({\it i})}}})}~,~~
\label{phi-2n+m4}\\
\frac{1}{2} \{x_{a^i}^{\mbox{\tiny{({\it i})}}}, 
z_{s^i}^{\mbox{\tiny{({\it i})}}}\}_{\mbox{\tiny{PB}}}
&=& 
-\sum_{b^i=1}^{2n^i} \tilde{g}^{(m^i)}_{a^ib^i} 
\frac{\partial (\tilde{G}_{1}^{\mbox{\tiny{({\it i})}}}, 
\cdots, \tilde{G}_{s^i-1}^{\mbox{\tiny{({\it i})}}},
\tilde{G}_{s^i}^{\mbox{\tiny{({\it i})}}}, 
\tilde{G}_{s^i+1}^{\mbox{\tiny{({\it i})}}}, \cdots, 
\tilde{G}_{m^i}^{\mbox{\tiny{({\it i})}}})}
{\partial (z_{1}^{\mbox{\tiny{({\it i})}}}, \cdots, 
z_{s^i-1}^{\mbox{\tiny{({\it i})}}}, 
x_{b^i}^{\mbox{\tiny{({\it i})}}}, 
z_{s^i+1}^{\mbox{\tiny{({\it i})}}}, \cdots,
z_{m^i}^{\mbox{\tiny{({\it i})}}})}~,~~
\label{phi-2n+m5}\\
\{z_{s^i}^{\mbox{\tiny{({\it i})}}}, 
z_{t^i}^{\mbox{\tiny{({\it i})}}}\}_{\mbox{\tiny{PB}}}&=& \nonumber \\
&&\!\!\!\!\!\!\!\!\!\!\!\!\!\!\!\!\!\!\!\!\!\!\!\!\!\!\!\!\!\!\!\!
\sum_{a^i,b^i=1}^{2n^i}\tilde{g}^{(m^i)}_{a^ib^i} 
\frac{\partial (\tilde{G}_{1}^{\mbox{\tiny{({\it i})}}}, 
\cdots, \tilde{G}_{s^i-1}^{\mbox{\tiny{({\it i})}}},
\tilde{G}_{s^i}^{\mbox{\tiny{({\it i})}}}, 
\tilde{G}_{s^i+1}^{\mbox{\tiny{({\it i})}}}, \cdots, 
\tilde{G}_{t^i-1}^{\mbox{\tiny{({\it i})}}},
\tilde{G}_{t^i}^{\mbox{\tiny{({\it i})}}}, 
\tilde{G}_{t^i+1}^{\mbox{\tiny{({\it i})}}}, \cdots,
\tilde{G}_{m^i}^{\mbox{\tiny{({\it i})}}})}
{\partial (z_{1}^{\mbox{\tiny{({\it i})}}}, \cdots, 
z_{s^i-1}^{\mbox{\tiny{({\it i})}}}, 
x_{a^i}^{\mbox{\tiny{({\it i})}}}, 
z_{s^i+1}^{\mbox{\tiny{({\it i})}}}, \cdots,
z_{t^i-1}^{\mbox{\tiny{({\it i})}}}, 
x_{b^i}^{\mbox{\tiny{({\it i})}}}, 
z_{t^i+1}^{\mbox{\tiny{({\it i})}}}, \cdots,
z_{m^i}^{\mbox{\tiny{({\it i})}}})}~,~~
\label{phi-2n+m6}
\end{eqnarray}
where $s^i< t^i$. 
We refer to the systems where the master equations 
are given by Eq. (\ref{N-eqf(m)}) or Eq. (\ref{N-eqf(m)2}) 
as generalized Nambu systems.

\section{Nambu systems in constrained Hamiltonian systems}
\label{Nambu systems in constrained Hamiltonian systems}
\subsection{Subject}
\label{Subject}

In the previous section, we found that a Hamiltonian system can be formulated 
as a Nambu system with multiplets including composite variables of $q$ and $p$.
The main feature of our formulation is the existence of induced constraints
that are required just for consistency between the variables.
Together with the Hamiltonian of the original system,
the induced constraints serve as Hamiltonians of the Nambu system. 
Therefore it is intriguing to study how constrained Hamiltonian systems,
systems with {\it physical} constraints, are cast into Nambu systems
in our formulation.

The relations between Nambu systems and constrained Hamiltonian systems
have been investigated by many authors \cite{BF,CK,R,MS,KT,KT2,LJ}.
To clarify the difference between previous works and our approach, 
here we give a brief summary of the results obtained so far.
In most works, Nambu systems are treated as the original systems, 
and studies have been carried out to find 
appropriate constrained Hamiltonian systems 
into which the Nambu systems can be embedded \cite{BF,CK,R,MS,KT,KT2}.
Specifically, it has been shown that 
Nambu equations (Eq. (\ref{N-eq})) are compatible with
the following equations:
\begin{eqnarray}
&& p_i = H_1 \frac{\partial H_2}{\partial x_i}~,\label{N-eq-con1}\\
&& \sum_{i=1}^3 \frac{\partial(H_1, H_2)}{\partial(x_i, x_j)}\frac{dx_i}{dt} = 0~.
\label{N-eq-con2}
\end{eqnarray}
Here $p_i$ ($i=1,2,3$) are the canonical conjugate momenta defined as 
$p_i \equiv \partial L/\partial \dot{x}_i$ with the Lagrangian
\begin{eqnarray}
L = H_1 \sum_{i=1}^3 \frac{\partial H_2}{\partial x_i}\frac{dx_i}{dt} ~.
\label{N-eq-conL}
\end{eqnarray} 
Equation (\ref{N-eq-con2}) can be derived as the Euler--Lagrange equation
from this Lagrangian,
and Eq. (\ref{N-eq-con1}) leads to
the relations $\phi_i \equiv p_i - H_1 {\partial H_2}/{\partial x_i} = 0$,
which can be regarded as constraints.
In this way, Nambu systems can be interpreted as Hamiltonian systems
with specific constraints.

On the other hand, researchers have studied whether constrained systems 
can be described as Nambu systems or not.
Specifically, it has been shown 
that constrained Hamiltonian systems can be formulated 
in terms of (a generalized form of) Nambu mechanics 
by introducing an extra phase-space variable \cite{LJ}.
For a system with canonical variables 
$(q_k, p_k)$ $(k=1, \cdots, n)$ and 
$m$ first class constraints $\phi_l(q_1, \cdots, p_n)=0$,
the equations of motion are given by
\begin{eqnarray}
\frac{d q_k}{d t} &=& \frac{\partial H}{\partial p_k} 
+ \sum_{l=1}^m \left(\
\frac{\partial \lambda_l}{\partial p_k}\phi_l
+ \lambda_l \frac{\partial \phi_l}{\partial p_k}
\right)~,\label{eq-constraints1}\\~~
\frac{d p_k}{d t} &=& -\frac{\partial H}{\partial q_k} 
- \sum_{l=1}^m \left(\
\frac{\partial \lambda_l}{\partial q_k}\phi_l
+ \lambda_l \frac{\partial \phi_l}{\partial q_k}
\right)~,
\label{eq-constraints2}
\end{eqnarray}
where $\lambda_l$ are Lagrange multipliers.
Equations (\ref{eq-constraints1}) and (\ref{eq-constraints2})
are derived from (a generalized form of) the Nambu equation
\begin{eqnarray}
\frac{d f}{dt} = \sum_{k=1}^n \frac{\partial (f, H_1, H_2)}{\partial (q_k, p_k, r)}~,
\label{N-eq-constraints}
\end{eqnarray}
where $f=f(q_1, \cdots, p_n)$, 
$r$ is an extra phase-space variable, 
and Hamiltonians are defined as  
\begin{eqnarray}
H_1 = H - r~,~~ H_2 = r+ \sum_{l=1}^m \lambda_l \phi_l~.
\label{H1H2-constraints}
\end{eqnarray}
The equation for $r$ is given by
\begin{eqnarray}
\frac{d r}{dt}= - \sum_{l=1}^m 
\left(\lambda_l \{\phi_l, H\}_{\mbox{\tiny{PB}}}
+\phi_l \{\lambda_l, H\}_{\mbox{\tiny{PB}}}
\right)
= - \sum_{l=1}^m \lambda_l \frac{d \phi_l}{dt},
\label{r-eq-constraints}
\end{eqnarray}
where the last equality holds after imposing constraints.
Requiring the extra variable $r$ to decouple from the dynamics,
i.e., $dr/dt = 0$,
we obtain ${d \phi_l}/{dt} = 0$.
 
Our approach differs from these previous works.  
Our starting point is not Nambu systems but Hamiltonian systems with constraints,
and we do not introduce extra variables but use redundant variables. 

\subsection{Nambu structure in constrained Hamiltonian systems}
\label{Nambu structure in constrained Hamiltonian systems}

Here we demonstrate that systems with first class constraints can be 
formulated as Nambu systems or generalized Nambu systems, 
without introducing extra degrees of freedom.

As a warm-up, 
we consider a system of two canonical doublets $(q_1, p_1)$ and $(q_2, p_2)$
with one first class constraint $\phi(q_1, p_1, q_2, p_2) = 0$ 
that is time independent:
$d\phi /dt = \{\phi, h\}_{\mbox{\tiny{PB}}}=0$.
Here $h=h(q_1, p_1, q_2, p_2)$ is the Hamiltonian of this system.
The constraint $\phi$ is associated with gauge degrees of freedom, 
and an auxiliary condition $\chi(q_1, p_1, q_2, p_2) = 0$ 
such that $\{\phi, \chi\}_{\mbox{\tiny{PB}}} \ne 0$
should be imposed to fix the freedom.

By an appropriate canonical transformation 
$(q_1, p_1, q_2, p_2) \to (Q_1, P_1, Q_2, P_2)$,
we can eliminate one of the canonical variables.
Here we show the case in which $P_2$ is eliminated as follows:\footnote{
The canonical transformation generated by the generator $G = \lambda \phi$
is the gauge transformation, and the infinitesimal one is given by
$\delta q_r = ({\partial{G}}/{\partial p_r}) \delta \xi$ and 
$\delta p_r = -({\partial{G}}/{\partial q_r}) \delta \xi$ $(r=1, 2)$.
Here $\lambda$ is an arbitrary function of the canonical variables and $\delta \xi$
is an infinitesimal parameter.
If we take $\lambda = (p_2 - \chi)/(({\partial{\phi}}/{\partial q_2})\xi)$
using a finite parameter $\xi$,
$p_2$ is transformed into $P_2 = \chi$ under the constraint $\phi=0$.
In the same way, one of the canonical variables can be eliminated
by an appropriate canonical transformation.
The variable to be eliminated depends on the physical systems.
}
\begin{eqnarray}
P_2 = \chi(q_1, p_1, q_2, p_2) = 0~.
\label{P2}
\end{eqnarray}
The new Hamiltonian $K$ is given by $K(Q_1, P_1, Q_2, P_2)=h(q_1, p_1, q_2, p_2)$,
and the original constraint $\phi$ is transformed as 
$\Phi(Q_1, P_1, Q_2, P_2)=\phi(q_1, p_1, q_2, p_2)$.
From $\{\phi, \chi\}_{\mbox{\tiny{PB}}} = \partial \Phi/\partial Q_2 \ne 0$, 
the constraint $\Phi=0$ can be solved by $Q_2$ to give $Q_2=Q_2(Q_1,P_1)$.
Then we obtain a constraint $\Psi \equiv Q_2 - Q_2(Q_1,P_1)=0$,
which is equivalent to the original constraint $\phi=0$.

If we consider a system described by the variables $(Q_1, P_1, Q_2)$
with the constraint $\Psi =0$,
it is easy to show that the equation of motion for any function $f=f(Q_1, P_1, Q_2)$
can be written in the form of the Nambu equation, 
\begin{eqnarray}
\frac{df}{dt} = \frac{\partial(f, H, \Psi)}{\partial(Q_1, P_1, Q_2)}~,
\label{N-eqQP}
\end{eqnarray}
where $H=H(Q_1, P_1, Q_2)=K(Q_1, P_1, Q_2, P_2=0)$ is the Hamiltonian.
In fact, for $f=Q_1$ and $f=P_1$, Hamilton's canonical equations of motion
\begin{eqnarray}
\frac{dQ_1}{dt} = \frac{\partial H(Q_1, P_1, Q_2(Q_1, P_1))}{\partial P_1} ~,
~~\frac{dP_1}{dt} = -\frac{\partial H(Q_1, P_1, Q_2(Q_1, P_1))}{\partial Q_1} ~,
\label{H-eqQP}
\end{eqnarray}
are derived from Eq. (\ref{N-eqQP}),
and for $f=\Psi$, we obtain 
time independence of the constraint, $d\Psi/dt = 0$.
On the other hand, for $f=Q_2$, the following equation is derived:
\begin{eqnarray}
\frac{dQ_2}{dt} = \frac{\partial(H(Q_1, P_1, Q_2), {\Psi})}{\partial(Q_1, P_1)}
= \frac{\partial(Q_2(Q_1, P_1), H(Q_1, P_1, Q_2(Q_1, P_1)))}{\partial(Q_1, P_1)}~.
\label{N-eqQ2}
\end{eqnarray}
Using $d\Psi/dt = 0$ and Eq. (\ref{N-eqQ2}),
we obtain Hamilton's equation of motion for $Q_2(Q_1, P_1)$,
\begin{eqnarray}
\frac{d Q_2(Q_1, P_1)}{d t} = 
\frac{\partial(Q_2(Q_1, P_1), H(Q_1, P_1, Q_2(Q_1, P_1)))}{\partial(Q_1, P_1)}~.
\label{H-eqQ2}
\end{eqnarray}

By referring to the results in Sect. 2.2,
let us formulate this system
by means of the composite triplet $X=X(Q_1, P_1)$, $Y=Y(Q_1, P_1)$, and $Z=Z(Q_1, P_1)$,
imposing a constraint $\tilde{G}(X, Y, Z)$ 
that is equivalent to the original constraint $\phi$.
We assume that $\partial(X, Y)/\partial(Q_1, P_1) \ne 0$, i.e.,
$Z$ is a redundant variable such that $Z=Z(X, Y)$.
If the variables satisfy the conditions
\begin{eqnarray}
\frac{\partial \tilde{G}}{\partial Z} = \frac{\partial(X,Y)}{\partial(Q_1,P_1)}~,~~
\frac{\partial \tilde{G}}{\partial X} = \frac{\partial(Y,Z)}{\partial(Q_1,P_1)}~,~~
\frac{\partial \tilde{G}}{\partial Y} = \frac{\partial(Z,X)}{\partial(Q_1,P_1)}~,
\label{dphi/dZ}
\end{eqnarray}
then the time evolution of any function $\tilde{f}=\tilde{f}(X,Y,Z)$ 
is given by the Nambu equation,
\begin{eqnarray}
\frac{d\tilde{f}}{dt} = 
\frac{\partial(\tilde{f}, \tilde{H}, \tilde{G})}{\partial(X, Y, Z)}~,
\label{N-eqXYZ}
\end{eqnarray}
where $\tilde{H}$ is equal to the original Hamiltonian, 
$\tilde{H}(X,Y,Z)=H(Q_1, P_1, Q_2)$.
We can define various types of Nambu systems depending on the choice of variables
and the constraint.

Here we present two simple examples.
First, if we choose 
$X=Q_1$, $Y=P_1$, $Z=Q_2(Q_1,P_1)$, and $\tilde{G}(X,Y,Z)=\Psi(Q_1, P_1, Q_2)$,
Eq. (\ref{N-eqXYZ}) clearly holds from Eq. (\ref{N-eqQP}).
Next, let us choose  
$Y=P_1$, $Z=Q_2(Q_1,P_1)$, and $\tilde{G}(X,Y,Z)=\Phi(Q_1, P_1, Q_2)$.
In this case, if the variable $X$ is given by
\begin{eqnarray}
X = \int \frac{\partial \tilde{G}}{\partial Z}dQ_1 
= \int \frac{\partial {\Phi}}{\partial Q_2}dQ_1~,~~
\label{XZ}
\end{eqnarray}
then the variables satisfy the conditions (Eq. (\ref{dphi/dZ})),
and the system is described as a Nambu system.

It is straightforward to extend the above $\lq\lq$warm-up" discussion 
to many degrees of freedom systems.
Consider a system of $n$ sets of canonical doublets $(q_k, p_k)$ $(k=1, \cdots, n)$
with $m$ kinds of first class constraints 
$\phi_s(q_1, p_1, \cdots, q_n, p_n) = 0$ $(s=1, \cdots, m)$.
The Hamiltonian of this system is given by $h(q_1, p_1, \cdots, q_n, p_n)$.
To fix the gauge degrees of freedom, 
$m$ kinds of auxiliary conditions 
$\chi_t(q_1, p_1, \cdots, q_n, p_n) = 0$ $(t=1, \cdots, m)$
that satisfy $\det \{\phi_s, \chi_t\}_{\mbox{\tiny{PB}}} \ne 0$
should be imposed.

By an appropriate canonical transformation 
$(q_1, p_1, \cdots, q_n, p_n) \to (Q_1, P_1, \cdots, Q_n, P_n)$,
we can eliminate some of the canonical variables.
Here we show the case in which $P_{n-m+t}$ are eliminated as follows:
\begin{eqnarray}
P_{n-m+t} = \chi_t(q_1, p_1, \cdots, q_n, p_n) = 0~.
\label{Ps}
\end{eqnarray}
The new Hamiltonian is given by 
$K=K(Q_1, P_1, \cdots, Q_n, P_n)=h(q_1, p_1, \cdots, q_n, p_n)$,
and the original constraints $\phi_s$ are transformed as 
$\Phi_s(Q_1, P_1, \cdots, Q_n, P_n)=\phi_s(q_1, p_1, \cdots, q_n, p_n)$. 
From $\det \{\phi_s, \chi_t\}_{\mbox{\tiny{PB}}} = 
\det (\partial \Phi_s/\partial Q_{n-m+t}) \ne 0$, 
the constraints $\Phi_{s}=0$ can be solved by $Q_{n-m+t}$ 
to give $Q_{n-m+t} = Q_{n-m+t}(Q_1,P_1, \cdots, Q_{n-m}, P_{n-m})$. 

By referring to the results in Sect. 2.4,
let us formulate this system by means of composite variables 
$X_{a}=X_{a}(Q_1, P_1, \cdots, Q_{n-m}, P_{n-m})$ $(a = 1, \cdots, 2n-2m)$
and $Z_s=Z_s(Q_1, P_1, \cdots, Q_{n-m}, P_{n-m})$,
imposing the constraints 
$\tilde{G}_s(X_{1}, \cdots, X_{2n-2m}, Z_{1}, \cdots, Z_{m})$ 
that are equivalent to the original constraints $\phi_s$.  
We assume that 
$\det\{X_a, X_b\}^{\prime}_{\mbox{\tiny{PB}}} \ne 0$, 
where the Poisson bracket is defined as
\begin{eqnarray}
\{A, B\}^{\prime}_{\mbox{\tiny{PB}}} 
\equiv \sum_{\alpha=1}^{n-m} 
\frac{\partial (A, B)}{\partial (Q_{\alpha}, P_{\alpha})}~.
\label{PBQPs}
\end{eqnarray}
This means that $Z_s$ are redundant variables such that 
$Z_s=Z_s(X_1, \cdots, X_{2n-2m})$.

If $\tilde{G}_s$ and $\tilde{g}^{(m)}_{ab}$ satisfy the following relations:
\begin{eqnarray}
\frac{1}{2} \{X_a, X_b\}^{\prime}_{\mbox{\tiny{PB}}}
&=& 
\tilde{g}^{(m)}_{ab} \frac{\partial (\tilde{G}_1, \cdots, \tilde{G}_{m})}
{\partial (Z_{1}, \cdots, Z_{m})}~,~~
\label{dphi/dZm1} \\
\frac{1}{2}\{X_a, Z_{s}\}^{\prime}_{\mbox{\tiny{PB}}}
&=& 
-\sum_{b=1}^{2n-2m} \tilde{g}^{(m)}_{ab} 
\frac{\partial (\tilde{G}_1, \cdots, \tilde{G}_{s-1},
\tilde{G}_{s}, \tilde{G}_{s+1}, \cdots, \tilde{G}_{m})}
{\partial (Z_{1}, \cdots, Z_{s-1}, X_b, Z_{s+1}, \cdots, Z_{m})}~,~~
\label{dphi/dZm2} \\
\{Z_s, Z_t\}^{\prime}_{\mbox{\tiny{PB}}}
&=&\!\!\!\!\!\!
\sum_{a,b=1}^{2n-2m}\tilde{g}^{(m)}_{ab} 
\frac{\partial (\tilde{G}_1, \cdots, 
\tilde{G}_{s-1}, \tilde{G}_{s}, \tilde{G}_{s+1}, \cdots, 
\tilde{G}_{t-1}, \tilde{G}_{t}, \tilde{G}_{t+1}, \cdots, \tilde{G}_{m})}
{\partial (Z_{1}, \cdots, Z_{s-1}, X_{a}, Z_{s+1},
\cdots, Z_{t-1}, X_{b}, Z_{t+1}, \cdots, Z_{m})},
\label{dphi/dZm3}
\end{eqnarray}
where $s<t$, then the time evolution of any function 
$\tilde{f}=\tilde{f}(X_{1}, \cdots, X_{2n-2m}, Z_{1}, \cdots, Z_{m})$ 
can be written as
\begin{eqnarray}
\frac{d \tilde{f}}{dt}
= \sum_{a,b=1}^{2n-2m} \tilde{g}^{(m)}_{ab} 
\frac{\partial (\tilde{f}, \tilde{H}, \tilde{G}_1, \cdots, \tilde{G}_{m})}
{\partial (X_{a}, X_{b}, Z_{1}, \cdots, Z_{m})}~,
\label{N-eqQPs}
\end{eqnarray}
where $\tilde{H}$ is the Hamiltonian, 
\begin{eqnarray}
\tilde{H}(X_{1}, \cdots, X_{2n-2m}, Z_{1}, \cdots, Z_{m})&=&\nonumber\\
&&\!\!\!\!\!\!\!\!\!\!\!\!\!\!\!\!\!\!\!\!\!\!\!\!\!\!\!\!\!\!\!\!\!\!\!\!\!\!\!
\!\!\!\!\!\!\!\!\!\!\!\!\!\!\!\!\!\!\!\!\!\!\!\!\!\!\!\!\!\!\!\!\!\!\!\!\!\!\!
K(Q_1, P_1, \cdots, Q_{n-m}, P_{n-m}, Q_{n-m+1}, P_{n-m+1}=0,
\cdots, Q_{n}, P_{n}=0)~.
\label{Hamiltonian}
\end{eqnarray}
We can define various types of Nambu systems depending on the choice of variables
and the constraint.
For example, in the case in which we take 
$X_{n-m+\alpha}=P_{\alpha}$ $(\alpha=1, \cdots, n-m)$,
$Z_s=Q_{n-m+s}$, and $\tilde{G}_s=\Phi_s$,
if the variables $X_{\alpha}$ are 
functions of $Q_{\alpha}$ that are given by
\begin{eqnarray}
X_{\alpha} = X_{\alpha}(Q_{\alpha})&=& 
2 \int \tilde{g}^{(m)}_{\alpha~n-m+\alpha} 
\frac{\partial (\tilde{G}_1, \cdots, \tilde{G}_{m})}
{\partial (Z_{1}, \cdots, Z_{m})} dQ_{\alpha} 
\nonumber \\
&=& 2 \int {g}^{(m)}_{\alpha~n-m+\alpha} 
\frac{\partial ({\Phi}_1, \cdots, {\Phi}_{m})}
{\partial (Q_{n-m+1}, \cdots, Q_{n})} dQ_{\alpha}~,
\label{Xalpha}
\end{eqnarray}
then the variables satisfy the conditions 
(Eqs. (\ref{dphi/dZm1})--(\ref{dphi/dZm3})),
and the system is described as a Nambu system.

In this way,
{\it Hamiltonian systems with first class constraints 
can be described as Nambu systems where the master equations 
are Nambu equations or generalized ones.}
It is straightforward to formulate constrained Hamiltonian systems 
as Nambu systems where both the original constraints and the induced ones
serve as Hamiltonians.
Such systems can be realized 
by introducing many more redundant variables.

\subsection{Example: relativistic free particle}
\label{Example: relativistic free particle}
As an example, we consider a relativistic particle moving freely 
in 4-dimensional Minkowski space. 
The motion is expressed by the space-time 4-vector $q^{\mu}=q^{\mu}(\tau)$ 
and corresponding canonical momenta $p_{\mu}=p_{\mu}(\tau)$
$(\mu=0, 1, 2, 3)$, where $\tau$ is the proper time.
Here we use the metric tensor $\eta_{\mu\nu} = \mbox{diag}(1, -1, -1, -1)$.
The system has four sets of canonical pairs $(q^{\mu}, p_{\mu})$
and one first class constraint 
\begin{eqnarray}
\phi = p^{\mu} p_{\mu} - m^2 c^2 = 0~,
\label{p2}
\end{eqnarray}
where $m$ is the mass of the particle, $c$ is the speed of light,
and Einstein's summation convention is used.
We impose an auxiliary condition $\chi = q^0 - c \tau = 0$ to fix the gauge freedom.
Performing a canonical transformation
\begin{eqnarray}
&&q^0 \to Q^0=\chi,~~~ q^i \to Q^i=q^i~,\nonumber \\
&&p_0 \to P_0=p_0,~~~p_i \to P_i=p_i~,
\label{Can}
\end{eqnarray}
where $i=1,2,3$, we can eliminate $Q^0$,
and the system is described by three sets of canonical pairs $(Q^i, P_i)$
with the new Hamiltonian $K = -cP_{0}$.
The original constraint $\phi$ is transformed as 
\begin{eqnarray}
\phi \to \Phi = P^{\mu} P_{\mu} - m^2 c^2 = 0~,
\label{p2t}
\end{eqnarray}
which has an equivalent expression,
\begin{eqnarray}
\Psi = P_{0} + \sqrt{{\mbox{\boldmath $P$}}^2 + m^2 c^2}=0~,
\label{explicit}
\end{eqnarray} 
where ${\mbox{\boldmath $P$}}^2 = \sum_{i} P_i^2$.
Then Hamilton's equations of motion for $Q^i$ and $P_i$ are given by
\begin{eqnarray}
\frac{d Q^i}{dt} &=& \frac{\partial K}{\partial P_i} 
= \frac{cP_i}{\sqrt{{\mbox{\boldmath $P$}}^2 + m^2 c^2}}~,
\label{Hamilton-rel1}\\
\frac{d P_i}{dt} &=& -\frac{\partial K}{\partial Q^i} = 0 ~.~~
\label{Hamilton-rel2}
\end{eqnarray}
\par
Using the results in Sect. 3.2,
let us construct Nambu systems that are equivalent to this system.
The target equation is Eq. (\ref{N-eqQPs}) with $a,b = 1, \cdots, 6$ and $m = 1$.
Here we present three types of Nambu systems.
In each case the Hamiltonian is given by $\tilde{H}=K$.

~~\\
(a) First we consider the simplest construction,
\begin{eqnarray}
&~& X_i = Q^i~,~~ X_{i+3}=Y_i=P_i~,~~ Z=P_0~,~~
\nonumber\\
&~& \tilde{G}= \Psi = Z + \sqrt{{\mbox{\boldmath $Y$}}^2 + m^2 c^2}~,
\label{G-rel1}
\end{eqnarray}
where ${\mbox{\boldmath $Y$}}^2 = \sum_{i} Y_i^2$. 
From Eq. (\ref{dphi/dZm1}) we obtain
\begin{eqnarray}
\tilde{g}^{(1)}_{ab}=\frac{1}{2}(\delta_{a,b-3}-\delta_{a-3,b})~,
\label{gab-rel1}
\end{eqnarray}
and Eq. (\ref{N-eqQPs}) becomes
\begin{eqnarray}
\frac{d \tilde{f}}{dt}
= \sum_{i=1}^3 \frac{\partial (\tilde{f}, \tilde{H}, \tilde{G})}
{\partial (X_i, Y_i, Z)}~.
\label{N-eq-rel1}
\end{eqnarray}
This equation reduces to Hamilton's equations of motion 
(Eqs. (\ref{Hamilton-rel1})--(\ref{Hamilton-rel2}))
and the energy conservation law .

~~\\
(b) Next we consider a slightly different construction,
\begin{eqnarray}
&~& X_i = Q^i~,~~ X_{i+3}=Y_i=P_i~,~~ Z=P_0^2~,~~
\nonumber\\
&~& \tilde{G}= \Phi = Z - {\mbox{\boldmath $Y$}}^2 - m^2 c^2~.
\label{G-rel2}
\end{eqnarray}
In this case, $\tilde{g}^{(1)}_{ab}$ has the same form as Eq.  (\ref{gab-rel1}),
because in both cases (a) and (b), 
$\{X_i, X_{i+3}\}^{\prime}_{\mbox{\tiny{PB}}}=1$ and 
${\partial \tilde{G}}/{\partial Z}= 1$ holds.
Therefore the resulting equation is same as Eq. (\ref{N-eq-rel1}).

~~\\
(c) Finally we consider a case in which 
${\partial \tilde{G}}/{\partial Z}\ne 1$,
\begin{eqnarray}
&~& X_i = 2P_0 Q^i~,~~ X_{i+3}=Y_i=P_i~,~~ Z=P_0~,~~
\nonumber\\
&~& \tilde{G}= \Phi = Z^{2} - {\mbox{\boldmath $Y$}}^2 - m^2 c^2~.
\label{G-rel3}
\end{eqnarray}
From Eq. (\ref{dphi/dZm1}) 
each component of the factor $\tilde{g}^{(1)}_{ab}$ is determined as follows:
\begin{eqnarray}
&~& \tilde{g}^{(1)}_{ij}=-\frac{X_iY_j-X_jY_i}{2Z^2}~, \nonumber \\
&~& \tilde{g}^{(1)}_{il}=\frac{1}{2}\delta_{i+3,l}~,
~~\tilde{g}^{(1)}_{li}=-\frac{1}{2}\delta_{l,i+3}~, \nonumber \\
&~& \tilde{g}^{(1)}_{lm}=0~,
\label{g-rel3}
\end{eqnarray}
where $i,j = 1, 2, 3$ and $l,m = 4, 5, 6$.
Although this is different from Eq.  (\ref{gab-rel1}),
we obtain the same equation as Eq. (\ref{N-eq-rel1}) again.
This is because 
${\partial (\tilde{f}, \tilde{H}, \tilde{G})}/{\partial (X_i, X_j, Z)}=0$ 
holds in this case.

\section{Conclusions and future work}
\label{Conclusions and future work}

We have given a variant formulation of Hamiltonian systems
in terms of variables including redundant degrees of freedom.
By use of a non-canonical transformation that enlarges the phase space
from $(q, p)$ to $(x_1, x_2, \cdots, x_{N})$,
we can reveal the Nambu mechanical structure hidden in a Hamiltonian system.
The Hamiltonians required for the Nambu mechanical description are given by
the Hamiltonian of the original system and constraints induced
due to the consistency between the variables. 
Our formulation can be extended to many degrees of freedom systems
and systems with first class constraints. 
Generalized forms of Nambu equation (Eqs. (\ref{N-eqf(m)}) and (\ref{N-eqf(m)2})) 
are required in some cases. 
Our approach to constrained systems is different from
the preceding works \cite{BF,CK,R,MS,KT,KT2,LJ}, i.e.,
we treat Nambu mechanics as effective mechanics,
and we introduce not extra degrees of freedom
but redundant degrees of freedom.

Our formulation is not just a change of description,
but gives a new insight into the statistical or quantum mechanical 
treatment of Hamiltonian systems.
For example, the Nambu equation (Eq. (\ref{H-eqf(gN)})) 
could give a basis for a novel quantization scheme for a Hamiltonian system.
In the present work, the constraints 
$(\tilde{G}_{1}, \tilde{G}_{2}, \cdots, \tilde{G}_{N})$
are {\it unphysical} ones and they all are set to zero. 
However, if we give them some appropriate values, 
Eq. (\ref{H-eqf(gN)}) could provide semi-classical equations for
quantum-mechanical expectation values \cite{OVP,AH}.
The non-vanishing $\tilde{G}_{b}$ come from quantum fluctuations,
e.g., if we take 
$x=\langle\hat{q}\rangle$, $y=\langle\hat{p}\rangle$, 
and $z=\langle\hat{q}^{2}\rangle$,
then $\tilde{G}=z-x^2$ has a non-zero value in general.
The same argument holds for statistical-mechanical expectation values.
Therefore we expect that 
the Nambu equation (Eq. (\ref{H-eqf(gN)})) with non-vanishing $\tilde{G}_{b}$
could be a master equation for 
the statistical or quantum mechanics of Hamiltonian systems.
More detailed studies will be presented in a future publication,
and they might provide important clues for handling  
the statistical or quantum mechanics of Nambu systems 
\cite{Takh,Gaut,Chat,Dito1,Dito2,Hoppe,Awata,Minic,Kawamura,Curt1,Zach}.

\section*{Acknowledgements}
This work was supported in part by scientific grants 
from the Ministry of Education, Culture,
Sports, Science and Technology under Grant Nos.~22540272 and 21244036 (Y.K.).

\appendix
\section{Derivation of the Nambu equation from the least action principle}
 
Let us derive the Nambu equation (Eq. (\ref{H-eqf(gN)})) 
from a Hamiltonian system using the least action principle.
Here we show the case with $N=3$, where the Nambu equations are given
in the form of Eq. (\ref{N-eq}).

Our starting point is the action integral such that
\begin{eqnarray}
S = \int \left(p \frac{dq}{dt} - H(q, p)\right) dt
= \int \left(p(x,y,z) \frac{d}{dt}q(x,y,z) - \tilde{H}(x, y, z)\right) dt~,
\label{S}
\end{eqnarray}
where $x=x(q,p)$, $y=y(q,p)$, $z=z(q,p)$,
and we assume that the Hamiltonian $H$ can be expressed by
\begin{eqnarray}
H(q, p) = \tilde{H}(x, y, z)~.
\label{H}
\end{eqnarray}
As mentioned in Sect. 2.2,
in general, 
$q$, $p$, and $H$ are not uniquely determined as functions of $x$, $y$, and $z$.
Our goal is to obtain the equations of motion 
that hold independently of the expressions of $q$ and $p$.

Let us regard $x=x(t)$, $y=y(t)$, and $z=z(t)$ as independent variables,
and consider their variation 
$x \to x + \delta x$, $y \to y + \delta y$, and $z \to z + \delta z$.
Integrating by parts and ignoring the surface terms, 
the variation of $S$ can be written as
\begin{eqnarray}
\delta S &=& \int 
\left(\delta p \frac{dq}{dt} - \delta q \frac{dp}{dt} - \delta \tilde{H}\right) dt
\nonumber \\
&=& \int \left[\left(\frac{\partial p}{\partial x} \delta x
+ \frac{\partial p}{\partial y} \delta y
+ \frac{\partial p}{\partial z} \delta z\right)\frac{dq}{dt}
- \left(\frac{\partial q}{\partial x} \delta x
+ \frac{\partial q}{\partial y} \delta y
+ \frac{\partial q}{\partial z} \delta z\right)\frac{dp}{dt}\right.
\nonumber \\
&& ~~~~~ - \left.\left(\frac{\partial \tilde{H}}{\partial x} \delta x
+ \frac{\partial \tilde{H}}{\partial y} \delta y
+ \frac{\partial \tilde{H}}{\partial z} \delta z\right)\right] dt
\nonumber \\
&=& \int \left[\left(\frac{\partial p}{\partial x} \frac{dq}{dt}
- \frac{\partial q}{\partial x} \frac{dp}{dt} 
- \frac{\partial \tilde{H}}{\partial x}\right) \delta x
+ \left(\frac{\partial p}{\partial y} \frac{dq}{dt}
- \frac{\partial q}{\partial y} \frac{dp}{dt} 
- \frac{\partial \tilde{H}}{\partial y}\right) \delta y \right.
\nonumber \\
&& ~~~~~ \left.
+ \left(\frac{\partial p}{\partial z} \frac{dq}{dt}
- \frac{\partial q}{\partial z} \frac{dp}{dt} 
- \frac{\partial \tilde{H}}{\partial z}\right) \delta z \right] dt~.
\label{deltaS}
\end{eqnarray}
Imposing the least action principle $\delta S = 0$, 
we obtain the equations of motion
\begin{eqnarray}
- \frac{\partial(q, p)}{\partial(x, y)} \frac{dy}{dt} 
+ \frac{\partial(q, p)}{\partial(z, x)} \frac{dz}{dt}
&=& \frac{\partial \tilde{H}}{\partial x}~,
\nonumber\\
- \frac{\partial(q, p)}{\partial(y, z)} \frac{dz}{dt}
+ \frac{\partial(q, p)}{\partial(x, y)} \frac{dx}{dt} 
&=& \frac{\partial \tilde{H}}{\partial y}~,
\nonumber \\
- \frac{\partial(q, p)}{\partial(z, x)} \frac{dx}{dt} 
+ \frac{\partial(q, p)}{\partial(y,z)} \frac{dy}{dt}
&=& \frac{\partial \tilde{H}}{\partial z}~,
\label{eqs-xyz}
\end{eqnarray}
where we use the chain rule for the derivative of a function $\tilde{u}(x,y,z)$,
\begin{eqnarray}
\frac{d}{dt}\tilde{u}(x,y,z) = 
\frac{\partial \tilde{u}}{\partial x}\frac{dx}{dt} 
+ \frac{\partial \tilde{u}}{\partial y}\frac{dy}{dt} 
+ \frac{\partial \tilde{u}}{\partial z}\frac{dz}{dt}~.
\label{chain}
\end{eqnarray}

In the case in which $\partial(x, y)/\partial(q, p) \ne 0$, $q$ and $p$ 
can be expressed by $x$ and $y$,
and then the following equations are derived:
\begin{eqnarray}
\frac{dx}{dt}= 
\frac{\partial \tilde{H}}{\partial y} \frac{\partial(x, y)}{\partial(q, p)}~,~~
\frac{dy}{dt}= 
-\frac{\partial \tilde{H}}{\partial x} \frac{\partial(x, y)}{\partial(q, p)}~,~~
\label{eqs-xy}
\end{eqnarray}
using the relations
\begin{eqnarray}
\frac{\partial(q, p)}{\partial(x, y)} = 
\left(\frac{\partial(x, y)}{\partial(q, p)}\right)^{-1}~,~~
\frac{\partial(q, p)}{\partial(z, x)} = 0~,~~ 
\frac{\partial(q, p)}{\partial(y, z)} = 0~.
\label{qp-xy}
\end{eqnarray}
In the same way, for the case of $\partial(y, z)/\partial(q, p) \ne 0$, 
$q$ and $p$ are expressed by $y$ and $z$,
and then the following equations are derived:
\begin{eqnarray}
\frac{dy}{dt}= 
\frac{\partial \tilde{H}}{\partial z} \frac{\partial(y, z)}{\partial(q, p)}~,~~
\frac{dz}{dt}= 
-\frac{\partial \tilde{H}}{\partial y} \frac{\partial(y, z)}{\partial(q, p)}~,~~
\label{eqs-yz}
\end{eqnarray}
and for the case of $\partial(z, x)/\partial(q, p) \ne 0$, we obtain 
\begin{eqnarray}
\frac{dz}{dt}= 
\frac{\partial \tilde{H}}{\partial x} \frac{\partial(z, x)}{\partial(q, p)}~,~~
\frac{dx}{dt}= 
-\frac{\partial \tilde{H}}{\partial z} \frac{\partial(z, x)}{\partial(q, p)}~.
\label{eqs-zx}
\end{eqnarray}
Combining Eqs. (\ref{eqs-xy}), (\ref{eqs-yz}), and (\ref{eqs-zx}), 
we can write down a set of equations,
\begin{eqnarray}
\frac{dx}{dt}&=& 
\frac{\partial \tilde{H}}{\partial y} \frac{\partial(x, y)}{\partial(q, p)}
- \frac{\partial \tilde{H}}{\partial z} \frac{\partial(z, x)}{\partial(q, p)}~,~~
\label{eqs-xyz2x}\\
\frac{dy}{dt}&=& 
\frac{\partial \tilde{H}}{\partial z} \frac{\partial(y, z)}{\partial(q, p)}
-\frac{\partial \tilde{H}}{\partial x} \frac{\partial(x, y)}{\partial(q, p)}~,~~
\label{eqs-xyz2y}\\
\frac{dz}{dt}&=& 
\frac{\partial \tilde{H}}{\partial x} \frac{\partial(z, x)}{\partial(q, p)}
-\frac{\partial \tilde{H}}{\partial y} \frac{\partial(y, z)}{\partial(q, p)}~,
\label{eqs-xyz2z}
\end{eqnarray}
which is consistent with every expression of $q$ and $p$. 
For example, in the case in which $q=q(x,y)$ and $p=p(x,y)$
($\partial(x, y)/\partial(q, p) \ne 0$),
Eqs. (\ref{eqs-xyz2x})--(\ref{eqs-xyz2y}) are reduced to Eq. (\ref{eqs-xy}),
and Eq. (\ref{eqs-xyz2z}) is equivalent to Hamilton's equation of motion.

Introducing a function $\tilde{G}=\tilde{G}(x, y, z)$ that 
satisfies the conditions (\ref{xyzG}),
Eqs. (\ref{eqs-xyz2x})--(\ref{eqs-xyz2z}) are 
rewritten as Nambu equations in the form of Eq. (\ref{N-eq}), 
\begin{eqnarray}
\frac{d x}{dt} = \frac{\partial (\tilde{H}, \tilde{G})}{\partial (y, z)}~,~~ 
\frac{d y}{dt} = \frac{\partial (\tilde{H}, \tilde{G})}{\partial (z, x)}~,~~
\frac{d z}{dt} = \frac{\partial (\tilde{H}, \tilde{G})}{\partial (x, y)}~.
\label{N-eqS}
\end{eqnarray}
These equations hold independently of the expression of $q$ and $p$.

\section{Hidden Nambu systems in Nambu systems}

Let us formulate a Nambu mechanical system 
with an $N$-plet $x_i$ $(i=1, \cdots, N)$
using $N+r$ variables $y_j=y_j(x_1, \cdots, x_N)$ $(j=1, \cdots, N+r)$,
where $r$ is a positive integer. 
We assume that at least $r+1$ of 
$\{y_{j_1}, \cdots, y_{j_{N}}\}_{\mbox{\tiny{NB}}}$
($j_{1}, \cdots, j_{N}=1, \cdots, N+r$)
do not vanish,
where $\{y_{j_1}, \cdots, y_{j_{N}}\}_{\mbox{\tiny{NB}}}$ 
is the Nambu bracket defined by Eq. (\ref{NB-N}).
Then the equation for any function 
$\tilde{f}(y_1, \cdots, y_{N+r}) = f(x_1, \cdots, x_N)$
can be written as
\begin{eqnarray}
\frac{d \tilde{f}}{dt} &=& 
\frac{\partial(f, H_1, \cdots, H_{N-1})}{\partial(x_1, x_2, \cdots, x_{N})} 
\nonumber \\
&=& \frac{1}{N!} \sum_{j_1, j_2, \cdots, j_{N}=1}^{N+r} 
\frac{\partial (\tilde{f}, \tilde{H}_1, \cdots, \tilde{H}_{N-1})}
{\partial (y_{j_1}, y_{j_2}, \cdots, y_{j_{N}})}
\{y_{j_1}, y_{j_2}, \cdots, y_{j_{N}}\}_{\mbox{\tiny{NB}}}~,
\label{N-eq(N)}
\end{eqnarray}
where $\tilde{H}_a(y_1, \cdots, y_{N+r}) 
= H_a(x_1, \cdots, x_N)$ $(a=1, \cdots, N-1)$.

Introducing functions $\tilde{G}_c(y_1, \cdots, y_{N+r}) = G_c(x_1, \cdots, x_N)$
$(c=1, \cdots, r)$ that satisfy the relation
\begin{eqnarray}
\frac{1}{r!}\sum_{j_{N+1}, \cdots, j_{N+r}=1}^{N+r} 
\varepsilon_{j_1 j_2 \cdots j_N j_{N+1} \cdots j_{N+r}} 
\frac{\partial (\tilde{G}_1, \cdots, \tilde{G}_r)}
{\partial (y_{j_{N+1}}, \cdots, y_{j_{N+r}})}
=\{y_{j_1}, y_{j_2}, \cdots, y_{j_{N}}\}_{\mbox{\tiny{NB}}}~,
\label{yjG}
\end{eqnarray}
Eq. (\ref{N-eq(N)}) is rewritten in the form of the Nambu equation,
\begin{eqnarray}
\frac{d \tilde{f}}{dt} 
= \frac{\partial(\tilde{f}, \tilde{H}_1, \cdots, \tilde{H}_{N-1}, 
\tilde{G}_1, \cdots, \tilde{G}_r)}
{\partial(y_1, y_2, \cdots, y_N, y_{N+1}, \cdots, y_{N+r})}~,
\label{N-eqf(N+1)}
\end{eqnarray}
where we use the formula for any functions 
$\tilde{A}_j=\tilde{A}_j(y_1, \cdots, y_{N+r})$,
\begin{eqnarray}
\frac{\partial (\tilde{A}_1, \tilde{A}_2, \cdots, \tilde{A}_N, 
\tilde{A}_{N+1}, \cdots, \tilde{A}_{N+r})}
{\partial (y_1, y_2, \cdots, y_N, y_{N+1}, \cdots, y_{N+r})}
&=& \!\!\!
\frac{1}{N!~r!} 
\sum_{j_1, j_2, \cdots, j_{N}, j_{N+1}, \cdots, j_{N+r}=1}^{N+r} 
\varepsilon_{j_1 j_2 \cdots j_N j_{N+1} \cdots j_{N+r}}
\nonumber \\
&&\!\!\!\!\!\!\!
\times~ 
\frac{\partial (\tilde{A}_1, \tilde{A}_2, \cdots, \tilde{A}_N)}
{\partial (y_{j_1}, y_{j_2}, \cdots, y_{j_N})}
\frac{\partial (\tilde{A}_{N+1}, \cdots, \tilde{A}_{N+r})}
{\partial (y_{j_{N+1}}, \cdots, y_{j_{N+r}})}~.
\label{J(N)}
\end{eqnarray}
By the use of Eq. (\ref{yjG}), 
it can be shown that the Nambu bracket between  
$G_c(x_1, \cdots, x_N)=\tilde{G}_c(y_1, \cdots, y_{N+r})$
and the arbitrary functions $u_a(x_1, \cdots, x_N)=\tilde{u}_a(y_1, \cdots, y_{N+r})$ 
$(a=1, \cdots, N-1)$ vanishes such that
\begin{eqnarray}
\{G_c, u_1, \cdots, u_{N-1}\}_{\mbox{\tiny{NB}}}
&=& \frac{1}{N!} \sum_{j_1, j_2, \cdots, j_N=1}^{N+r} 
\frac{\partial (\tilde{G}_c, \tilde{u}_1, \cdots, \tilde{u}_{N-1})}
{\partial (y_{j_1}, y_{j_2}, \cdots, y_{j_{N}})} 
\{y_{j_1}, y_{j_2}, \cdots, y_{j_{N}}\}_{\mbox{\tiny{NB}}}
\nonumber \\
&=& \frac{1}{N!~r!} 
\sum_{j_1, j_2, \cdots, j_N, j_{N+1}, \cdots, j_{N+r}=1}^{N+r} 
\varepsilon_{j_1 j_2 \cdots j_N j_{N+1} \cdots j_{N+r}} 
\nonumber \\
&&~~~~~~~~~~~~~
\times~
\frac{\partial (\tilde{G}_c, \tilde{u}_1, \cdots, \tilde{u}_{N-1})}
{\partial (y_{j_1}, y_{j_2}, \cdots, y_{j_{N}})} 
\frac{\partial (\tilde{G}_{1}, \cdots, \tilde{G}_{r})}
{\partial (y_{j_{N+1}}, \cdots, y_{j_{N+r}})}
\nonumber \\
&=& \frac{\partial (\tilde{G}_c, \tilde{u}_1, \cdots, \tilde{u}_{N-1}, 
\tilde{G}_1, \cdots, \tilde{G}_r)}
{\partial (y_{1}, y_{2}, \cdots, y_{N}, y_{N+1}, \cdots, y_{N+r})}=0~.
\label{Gu=0-N}
\end{eqnarray}
Hence $G_c$ are constants.
We can eliminate the constants by redefining $G_c$, 
and the resulting $\tilde{G}_c(y_1,\cdots, y_{N+r})=0$ can be regarded 
as {\it induced constraints}, which are associated with enlarging the phase space
from $(x_1, \cdots, x_{N})$ to $(y_1, \cdots, y_{N+r})$.

In this way, 
{\it Nambu systems with an $N$-plet $x_i$ $(i=1, \cdots, N)$
can be formulated as Nambu systems with an $N+r$-plet 
$y_j=y_j(x_1, \cdots, x_N)$ $(j=1, \cdots, N+r)$.}


%
%

\end{document}